\documentclass[%
 reprint,
superscriptaddress,
showpacs,
 amsmath,amssymb,
 aps,
prmaterials,
floatfix, longbibliography ]{revtex4-2}
\usepackage{graphicx}
\usepackage{dcolumn}
\usepackage{bm}
\usepackage{soul,color}

\usepackage[colorlinks,linkcolor=blue,anchorcolor=blue,citecolor=blue,urlcolor=blue]{hyperref}
\usepackage[]{datetime}
\begin{document}

\title{
       Thermal decomposition of hydrated graphite oxide:
       A computational study}

\author{Andrii Kyrylchuk}
\affiliation{Institute of Organic Chemistry,
             National Academy of Sciences of Ukraine,
             Murmanska Str.~5, 02660 Kyiv, Ukraine}

\author{Pranav Surabhi}
\affiliation{Indian Institute of Technology,
             Khandwa Road, Simrol,
             Indore 453552, India}

\author{David Tom\'anek}
\email[E-mail: ]{tomanek@msu.edu}%
\affiliation{Physics and Astronomy Department,
             Michigan State University,
             East Lansing, Michigan 48824, USA}

\date{\today}

\begin{abstract}
We study the behavior of hydrated graphite oxide (GO) at high
temperatures using thermally accelerated molecular dynamics
simulations based on {\em ab initio} density functional theory.
Our results suggest that GO, a viable candidate for water
treatment and desalination membranes, is more heat resilient than
currently used organic materials. The system we consider to
represent important aspects of thermal processes in highly
disordered GO is a hydrated GO bilayer in vacuum. Our study
provides microscopic insight into reactions involving water and
functional epoxy-O and OH-groups bonded to graphene layers, and
also describes the swelling of the structure by water vapor
pressure at elevated temperatures. We find the system to withstand
simulation temperatures up to ${\approx}2,500$~K before the
graphitic layers start decomposing, implying the possibility of
cleaning biofouling residue from a GO-based membrane by heating in
an inert gas atmosphere.
\end{abstract}


\maketitle
\renewcommand\thesubsection{\arabic{subsection}}



\section{Introduction}

Lack of potable water is agreeably the most urgent problem of
humankind today. Whereas water in oceans is plentiful, it requires
desalination prior to human consumption~\cite{Kucera}. The most
common desalination process is reverse osmosis
(RO)~\cite{Cohen2017}. The key component of an RO desalination
plant is a strong semi-permeable membrane that lets water
molecules pass, but rejects ions. Current membranes, based on
polyamide and other organic nanoporous substances, display
satisfactory ion rejection at acceptable water permeation
rates~\cite{Epsztein2020}, implying that new materials should
offer only marginal improvement in performance~\cite{Elimelech20}.
Yet these optimized membranes have serious limitations in their
mechanical, thermal and chemical stability~\cite{Werber2016}. An
urgent problem occurring in all of water treatment is the
formation of a biofouling residue that clogs the
membrane~\cite{{Kucera19},{Werber2016}}. Cleaning this residue
using chemical agents is of limited use for organic membranes,
since such chemicals also attack the membrane
material~\cite{{Porcelli2010},{Deemer19}}. The desalination
community has long been waiting for a paradigm shift that
alleviates this problem~\cite{{Elimelech2011},{Shahzad2019}}.

Here we present results of an {\em ab initio} density functional
theory (DFT) molecular dynamics (MD) simulation addressing
thermally driven structural changes in a bilayer of hydrated
graphite oxide (GO) in vacuum. This rather artificial system was
selected as a model to study microscopic details of the thermal
decomposition of GO, which has been demonstrated to allow water
permeation while rejecting solvated ions in the
feed~\cite{Boehm1961} when sandwiched in-between layers of carbon
nanotube buckypaper and carbon fabric for containment and
mechanical strength~\cite{DT274}. In order to observe slow
processes in the short time frame of the simulation, we
artificially raised the system temperature in our thermally
accelerated MD studies~\cite{MD-temperature}. At simulation
temperatures up to $4,000$~K, we found that water molecules in the
interlayer region are rather decoupled from the GO layers and only
marginally affect their behavior except for swelling the structure
by water vapor pressure. In presence of nearby water molecules,
some epoxy-O atoms move from their bridge to the on-top site,
turning into radicals and changing the configuration of the
connected carbon atom from $sp^3$ to $sp^2$. In a similar way, in
presence of water, hydrogen atoms often detach from adsorbed
hydroxy groups and turn them into epoxy groups. Both processes
facilitate buckling and local fracture of the graphitic backbone
above $2,500$~K. At higher temperatures, we observe the
destruction of the graphitic backbone itself.
Oxygens in the functional groups migrate from the faces to the
reactive exposed edges of the graphitic flakes, turning GO into
hydrophobic reduced GO (rGO) not subject to swelling.

As indicated above, the vast majority of state-of-the-art
desalination membranes use nanoporous organic compounds in the
active layer. Yet as early as 1961, Boehm
reported~\cite{Boehm1961} that
GO~\cite{{Brodie1859},{Hummers1958}}, an inorganic compound
related to graphite, is impermeable to liquids other than polar
water. The same study~\cite{Boehm1961} found that GO was
practically impermeable to anions, but permeable to cations. With
few notable exceptions~\cite{{Boretti2018},{Wang2021}}, the
reported selectivity and permeability by water were barely noticed
by the membrane community, since GO is essentially a disordered
powder that is hard to characterize and to contain.

GO has undergone serious characterization analysis by experimental
and theoretical techniques since its discovery and development of
effective synthesis techniques~\cite{{Brodie1859},{Hummers1958}}.
The basic conclusion was that GO partly resembles what we now know
as intercalated graphite, by containing finite-size graphene
flakes, functionalized mainly by O- and OH- groups, which are
largely disordered and separated by water
molecules~\cite{Hofmann1934}. Progress has also been made in
improving flake alignment within this layered
system~\cite{Akbari16}. Interestingly, presence of thermally
stable and electrically conductive graphitic material has been
observed to reduce biofouling in membranes~\cite{%
{Zhang2013},{Hu2016},{Huang2016},{Han2016},{Rashid2017},{Ai2018},%
{Seo2018},{Perez-Roa2009},{Lu2018},{Ho2018}}.

Whereas advanced experimental techniques have provided a
significant amount of structural and chemical information,
microscopic knowledge of atomic-level processes in GO is scarce
due to its complex structure and disorder that is changing in
presence of water. Probably the best description of swelling and
thermal reduction of GO to rGO to date is provided in the 1934
study by Hofmann~\cite{Hofmann1934}.

Computer simulations of this complex, disordered system are
seriously limited by the total simulation time and the size of the
system, where the quantum nature of interatomic interactions is
essential for a correct prediction of atomic-scale processes.
Thus, atomic-level studies of a substance as complex as hydrated
GO are only possible in a model system that is much simpler, yet
closely represents at least one of its aspects. Finding such an
ideal model system in nature is almost impossible. Theory, on the
other hand, deals with simplified model systems, primarily to
obtain an understanding of ongoing atomic-level processes. Such
studies then provide useful information only for a limited number
of phenomena observed in the more complex GO system.

Related theoretical studies have considered water flow in-between
GO layers~\cite{{DT285},{Elimelech19}} and inside carbon
nanotubes~\cite{{Corry2008},{DT287}}. Most important for selective
permeation by water and rejection of ions were model studies of
in-layer pores within GO monolayers~\cite{DT274}. A pore
consisting of a pair of in-layer GO edges was characterized by the
pore width, the edge type being either armchair or zigzag, and
termination by H- O- or OH- groups. Results of that study
indicated that selective permeation by water and rejection of ions
is possible for pores not narrower than $0.7$~nm and not wider
than $0.9$~nm, in agreement with the consensus in the water
treatment community~\cite{Epsztein2020}.

Thermal stability of GO and its disintegration at high
temperatures has been investigated theoretically in only few
studies~\cite{Paci2007}. Yet the behavior of GO at elevated
temperatures is the key to answering the question, whether GO
membranes may be cleaned by heating, possibly in an inert gas
atmosphere. An MD study of a model system at elevated temperatures
may reveal if OH- and O- functional groups, which lower the
mechanical strength of GO, remain attached up to the temperature,
when C-C bonds break, and thus induce the disintegration of the
graphitic backbone. An MD study should also reveal if interlayer
water turned vapor may exert sufficient pressure to separate
adjacent GO layers, and whether water molecules may exchange atoms
and modify the structure of functional groups.

Theory has the unique advantage to describe a hypothetical system
that should exist, but is nearly impossible to synthesize. The
single purpose of such a system is to study particular aspects of
behavior that would be hard to isolate in a more complex system.
The specific system we address in this study is an infinite,
defect-free bilayer of graphite oxide, suspended in vacuum, which
contains water molecules in the inter-layer region in much the
same way as the bulk system does. This geometry also offers the
benefit of not suppressing changes in the interlayer separation by
artificial boundary conditions.

\begin{figure}[t]
    \centering
    \includegraphics[width=0.8\columnwidth]{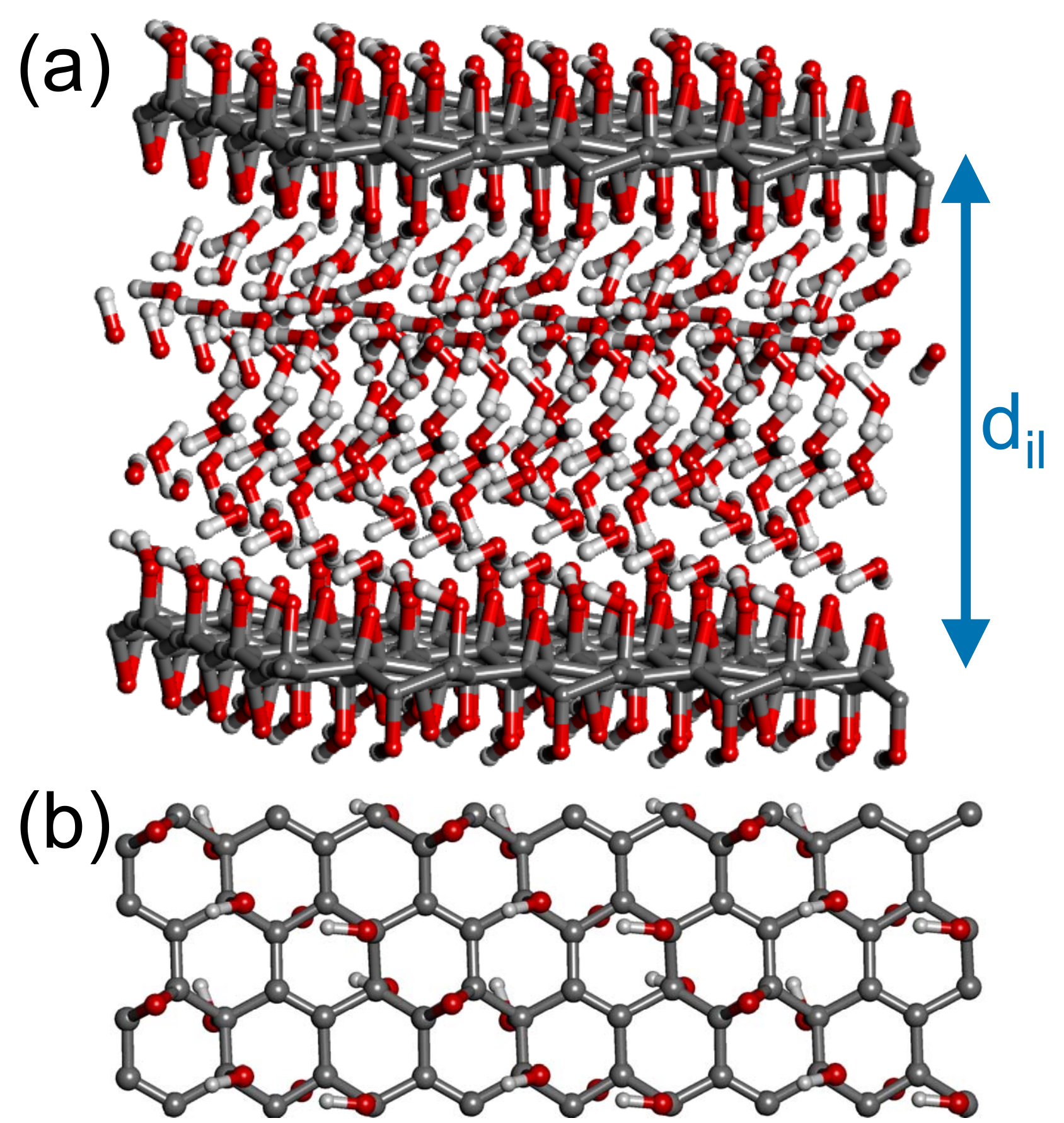}
    \caption{%
        Ball-and-stick models of a GO bilayer containing H$_2$O
        molecules in the interlayer region.
        (a) A $2{\times}2$ GO bilayer supercell in perspective
        view. $d_{il}$ is the interlayer distance.
        (b) The top layer in top view displaying O atoms
        forming 1,3-epoxy groups and OH groups chemisorbed
        on the graphene layer. }%
    \label{fig1}
\end{figure}

\section{Computational approach}


Our computational approach to study hydrated GO is based on
\textit{ab initio} density functional theory (DFT) as implemented
in the {\textsc{SIESTA}}~\cite{SIESTA} code. %
We used the nonlocal Perdew-Burke-Ernzerhof (PBE)~\cite{PBE}
exchange-correlation functional in the {\textsc{SIESTA}} code,
norm-conserving Troullier-Martins
pseudopotentials~\cite{Troullier91}, a double-$\zeta$ basis
including polarization orbitals, and a mesh cutoff energy of
$180$~Ry to determine the self-consistent charge density. The
hydrated bilayer of graphite oxide, described by the
Lerf-Klinowski model~\cite{Lerf1998}, is shown in Fig.~\ref{fig1}.
Of the 308 atoms in the unit cell, there are 96 carbon atoms in
the two graphitic layers and 44 H$_2$O molecules in the interlayer
region. The remaining 32 H and 48 O atoms form epoxy-O- and
OH-functional groups covalently bonded to the graphene layer. We
have used periodic boundary conditions throughout the study, with
replicas of the bilayer initially separated by a large distance of
$30$~{\AA}. While computationally rather demanding, the DFT-PBE
energy functional is free of adjustable parameters and has been
used extensively to provide an unbiased description of water and
its interaction with solids~\cite{{Cicero2008},{Ambrosetti2011}}.

\begin{figure}[t]
    \centering
    \includegraphics[width=0.9\columnwidth]{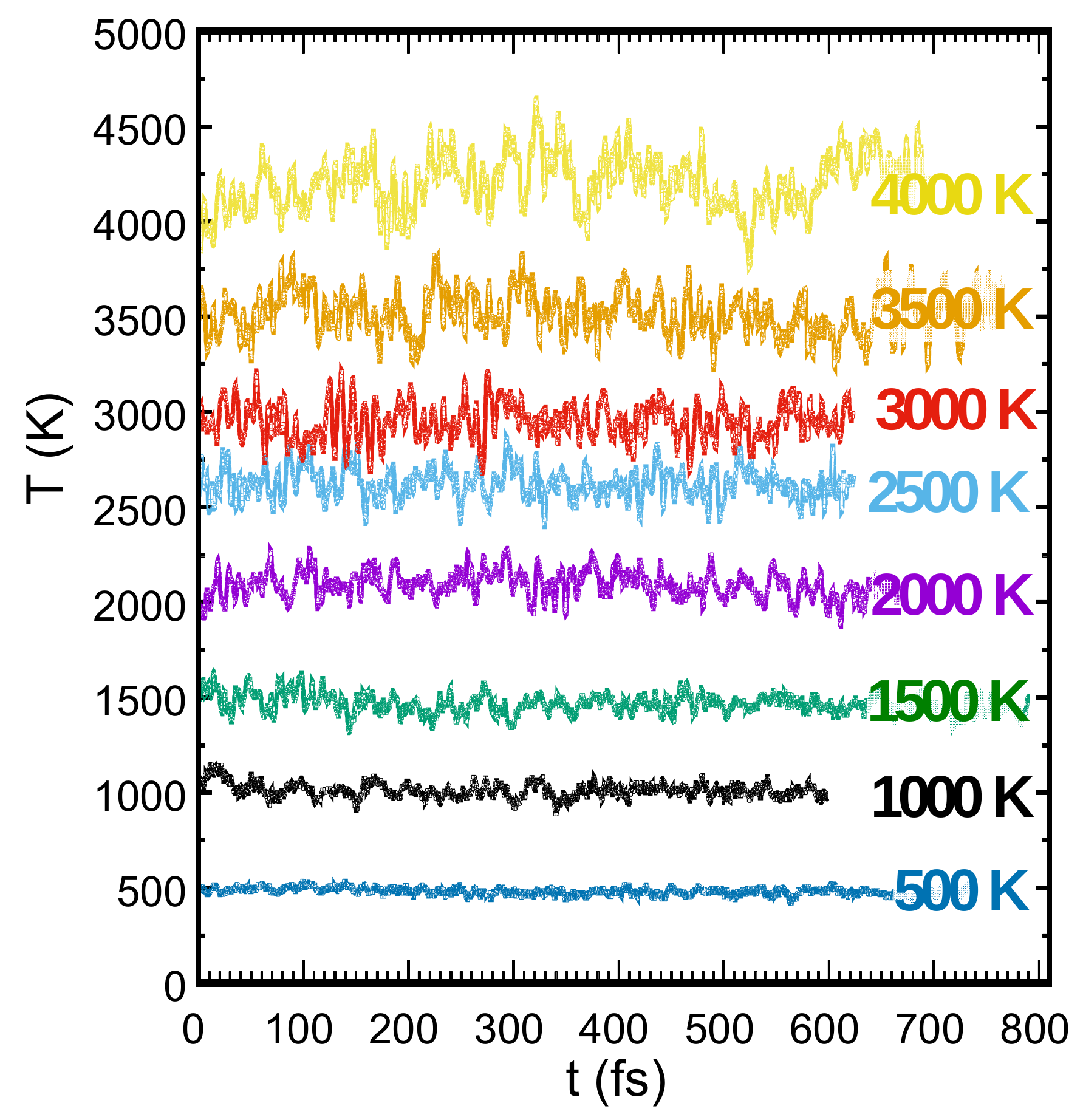}
    \caption{%
        Temperature fluctuations during an NVE simulation
        of the hydrated GO bilayer unit cell containing 308 atoms.
        The temperatures, at which each ensemble has been initialized,
        are indicated by the labels and are discussed in the text.}%
    \label{fig2}
\end{figure}

\begin{figure}[b]
    \centering
    \includegraphics[width=1.0\columnwidth]{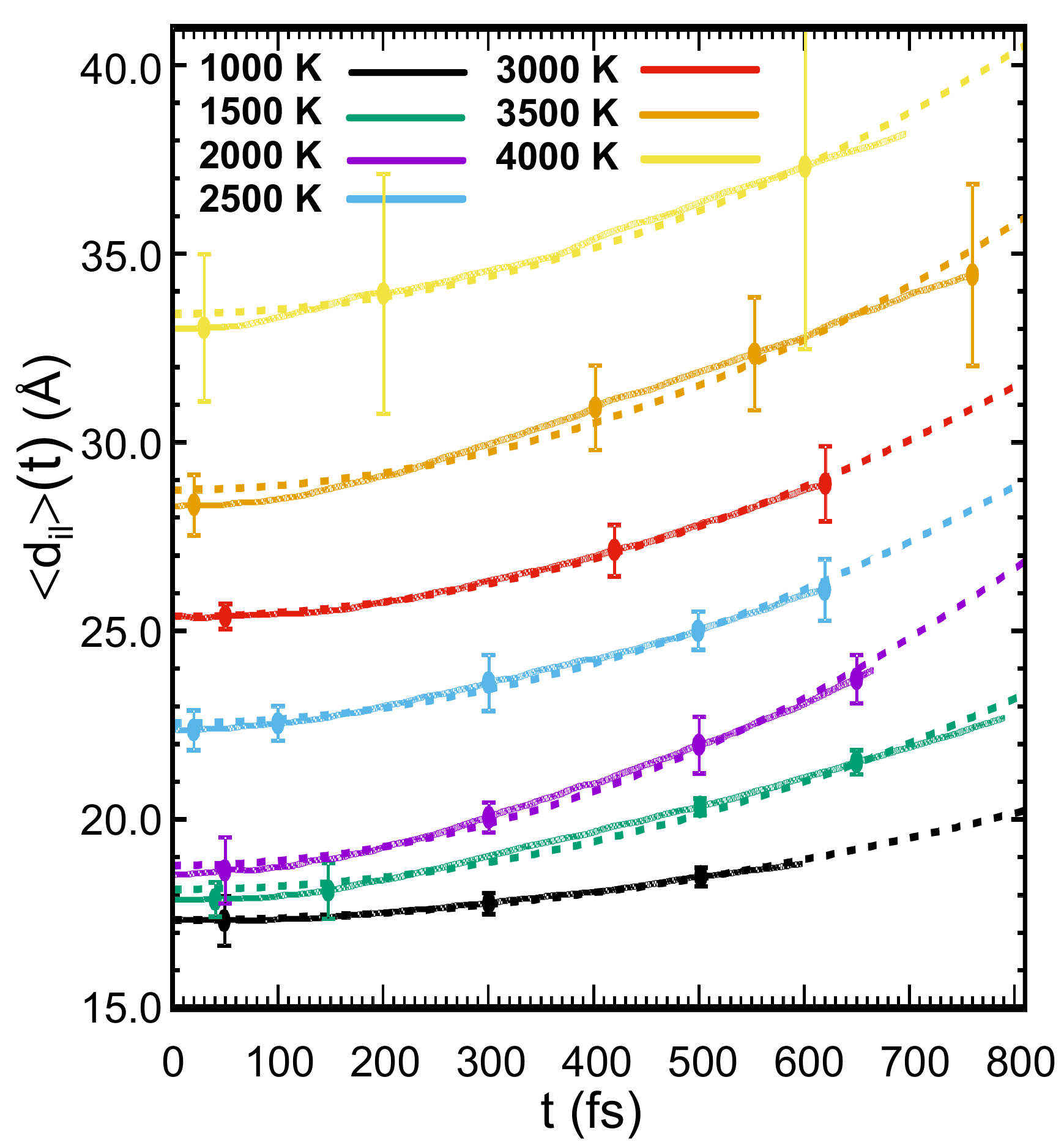}
    \caption{%
        The interlayer distance $d_{il}$, averaged across the unit cell,
        at different simulation temperatures $T$ as a function of
        time $t$.}%
    \label{fig3}
\end{figure}


Slow chemical processes pose a significant challenge to atomistic
MD simulations, which we use. Yet according to a well-known
chemical rule-of-thumb, the speed of a reaction doubles upon a
temperature increase by $10$~K. The high computational cost of the
{\em ab initio} force field and the large size of the unit cell
limit our calculations to ${\lesssim}1$~ps. To observe processes,
which occur in nature on the time scale of seconds, within the
short time period of $1$~ps, chemical processes have to be
accelerated by the factor of $10^{12}$. This may be achieved using
an approach we call thermally accelerated dynamics. Taking the
above chemical rule-of-thumb seriously, to at least one order of
magnitude, a desirable acceleration should occur upon raising the
temperature of the system artificially by
${\Delta}T{\approx}200-2,000$~K. Consequently, processes occurring
during $<1$~ps at the simulation temperature $T$ are expected to
occur at a much lower temperature $T-{\Delta}T$ in nature on the
time scale of seconds or longer.


We selected simulation temperatures in the range from $500$~K to
$4,000$~K to initially equilibrate the system for a time period of
$60$~fs by treating it as a canonical (NVT) ensemble regulated by
a Nos\'e thermostat. Following this equilibration period, the
system has been treated as a microcanonical (NVE) ensemble in
order to avoid artifacts caused by the thermostat. We found that
$0.3$~fs time steps were sufficiently short to keep the total
energy in the NVE ensemble conserved, while allowing the
temperature to fluctuate, as seen in Fig.~\ref{fig2}. As expected,
the temperature fluctuations increase with system temperature $T$.
In view of the relatively large unit cell size, the range of
temperature fluctuations in the NVE ensemble is adequate and
should not affect our conclusions.


\section{Results}

Our MD results depicting the behavior of a hydrated GO monolayer
at $T=500$~K are presented in Video~\ref{video1}. The reason for
GO being hydrophilic is that the calculated chemisorption energy
of $0.73$~eV of an isolated H$_2$O molecule on GO exceeds its
calculated hydration energy of $0.41$~eV. Still, this energy is
rather low, so that molecules surrounding the monolayer detach
easily and evaporate into the vacuum region above and below even
on the short time scale below $1$~ps at $500$~K.

As mentioned before, we consider a bilayer with water contained in
the interlayer region to keep GO hydrated. Since water molecules
detach easily above the top and below the bottom layers, they do
not affect the dynamics of the bilayer and will be omitted in our
simulations. In our periodic system, infinite hydrated bilayers
are separated by a substantial vacuum region that eliminates the
interaction between replicas even under the most extreme
conditions. In this geometry, all atoms in the bilayer are
completely free to move rather than being constrained in the
out-of-plane direction as they would be in bulk GO with imposed
periodicity in that direction.

\subsection{Atomic motion at elevated temperatures}

\begin{video}
\includegraphics[width=0.50\columnwidth]{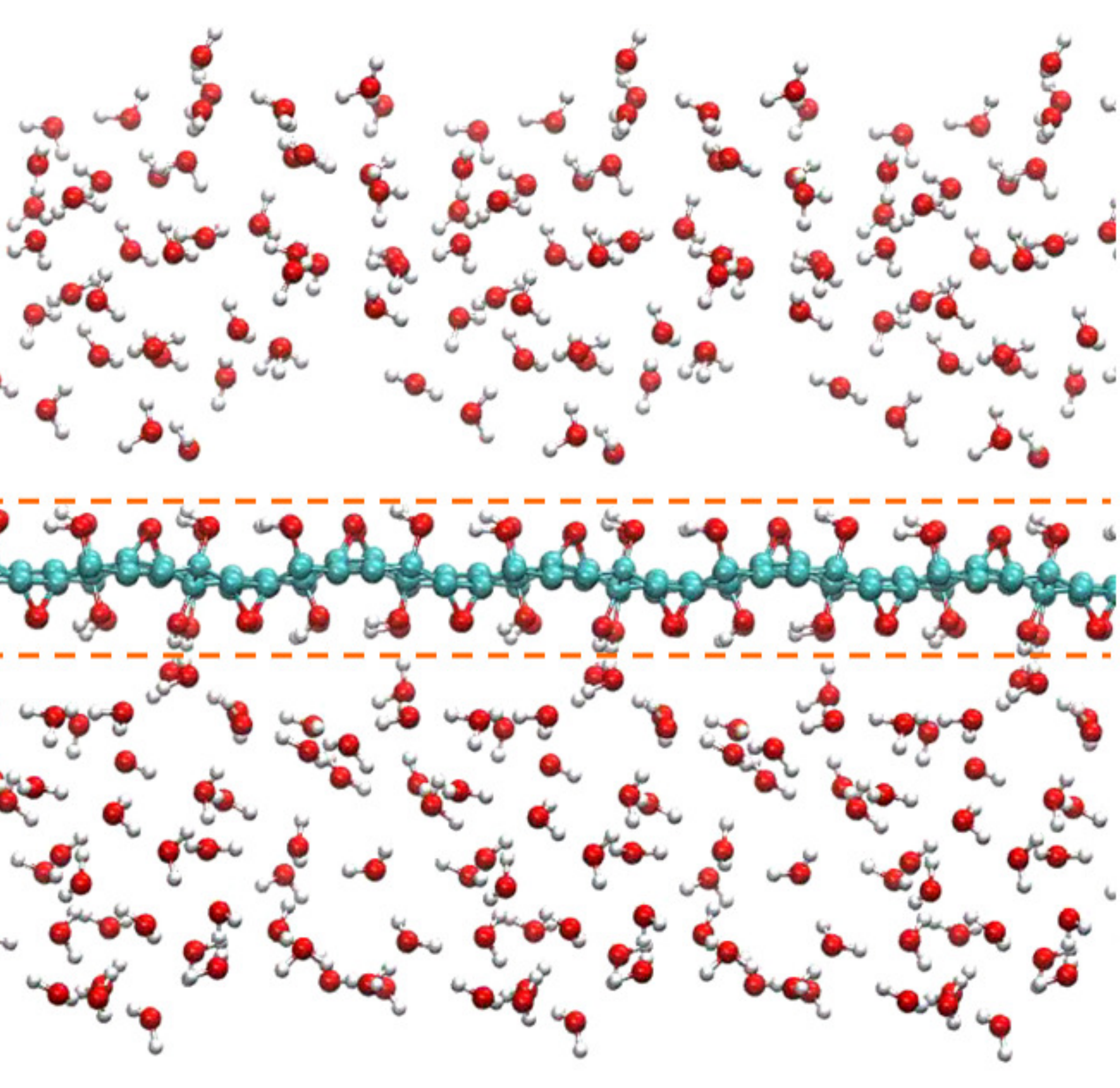}
\setfloatlink{https://nanoten.com/tomanek/manuscripts/DT290v1.mp4} %
\caption{Microcanonical MD simulation of a hydrated GO monolayer
that had been initially equilibrated at $T=500$~K, visualizing the
slow detachment of adsorbed water from both sides.
A $3{\times}3{\times}1$ supercell is shown for clarity. %
}
\label{video1}
\end{video}

\begin{video}[b]
\includegraphics[width=0.50\columnwidth]{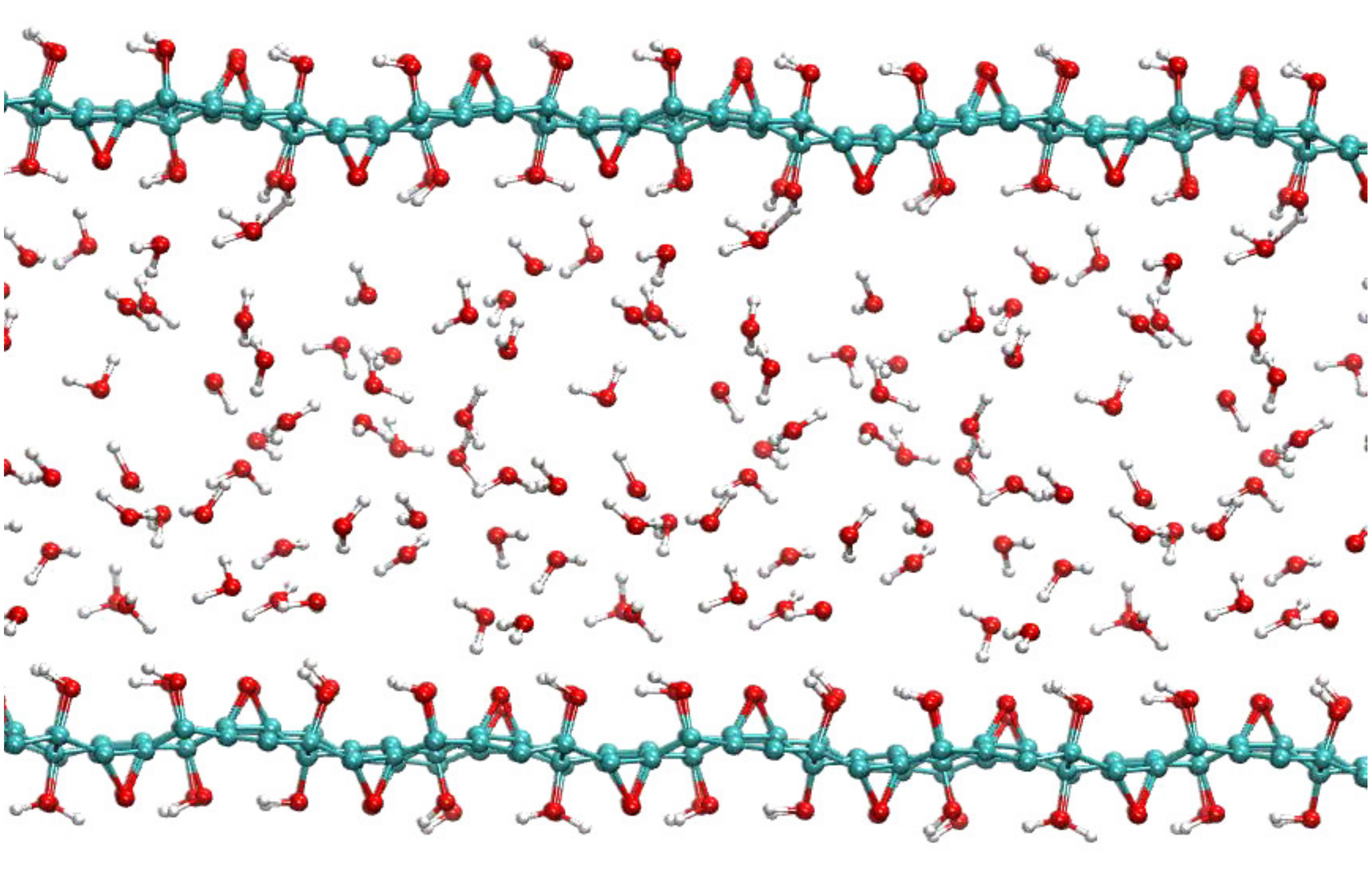}
\setfloatlink{https://nanoten.com/tomanek/manuscripts/DT290v2.mp4} %
\caption{Microcanonical MD simulation of a hydrated GO bilayer
that had been initially equilibrated at $T=500$~K.
A $3{\times}3{\times}1$ supercell is shown for clarity. %
}
\label{video2}
\end{video}

\begin{video}[h]
\includegraphics[width=0.40\columnwidth]{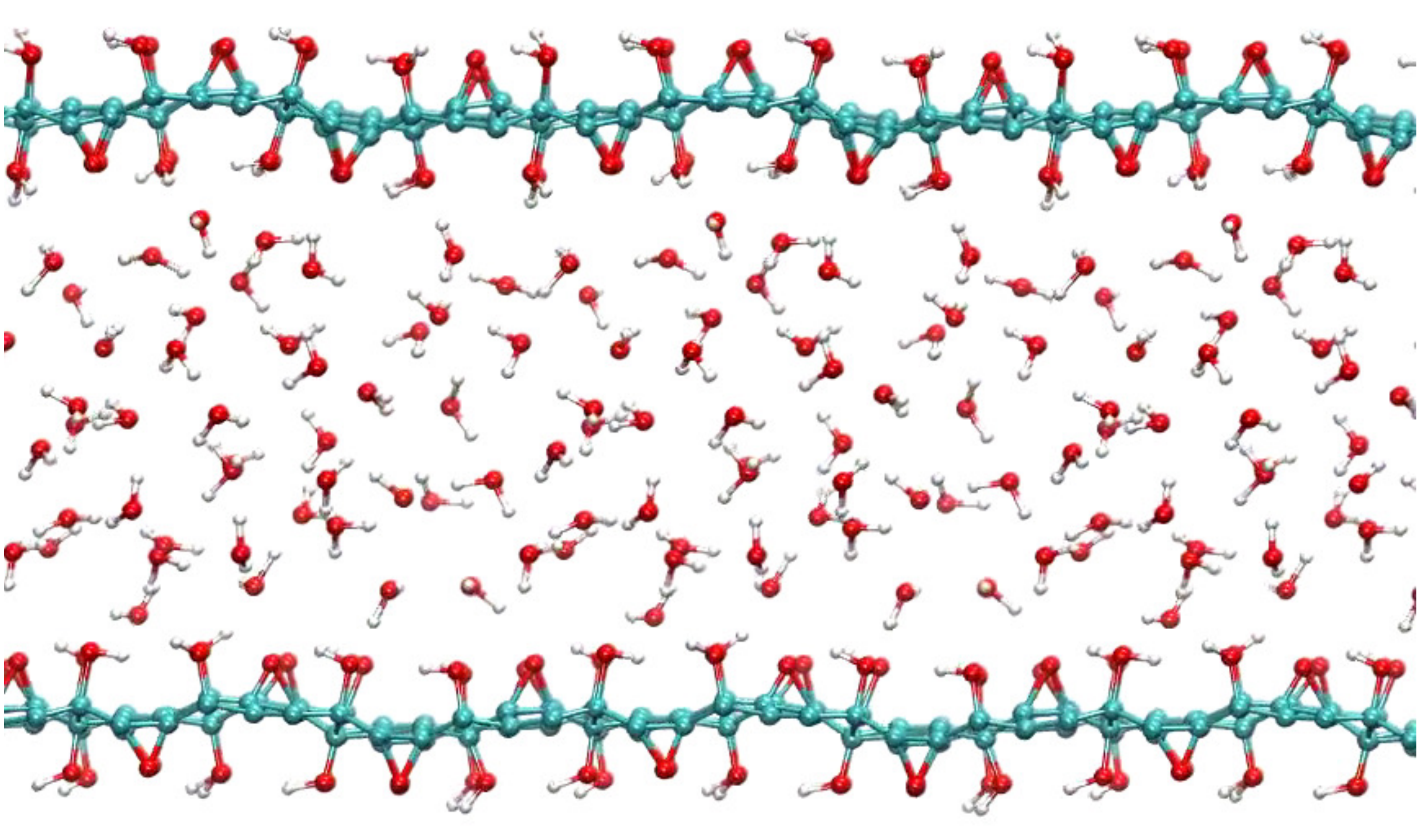}
\setfloatlink{https://nanoten.com/tomanek/manuscripts/DT290v3.mp4} %
\caption{Microcanonical MD simulation of a hydrated GO bilayer
that had been initially equilibrated at $T=1000$~K. %
}
\label{video3}
\end{video}

\begin{video}[h]
\includegraphics[width=0.50\columnwidth]{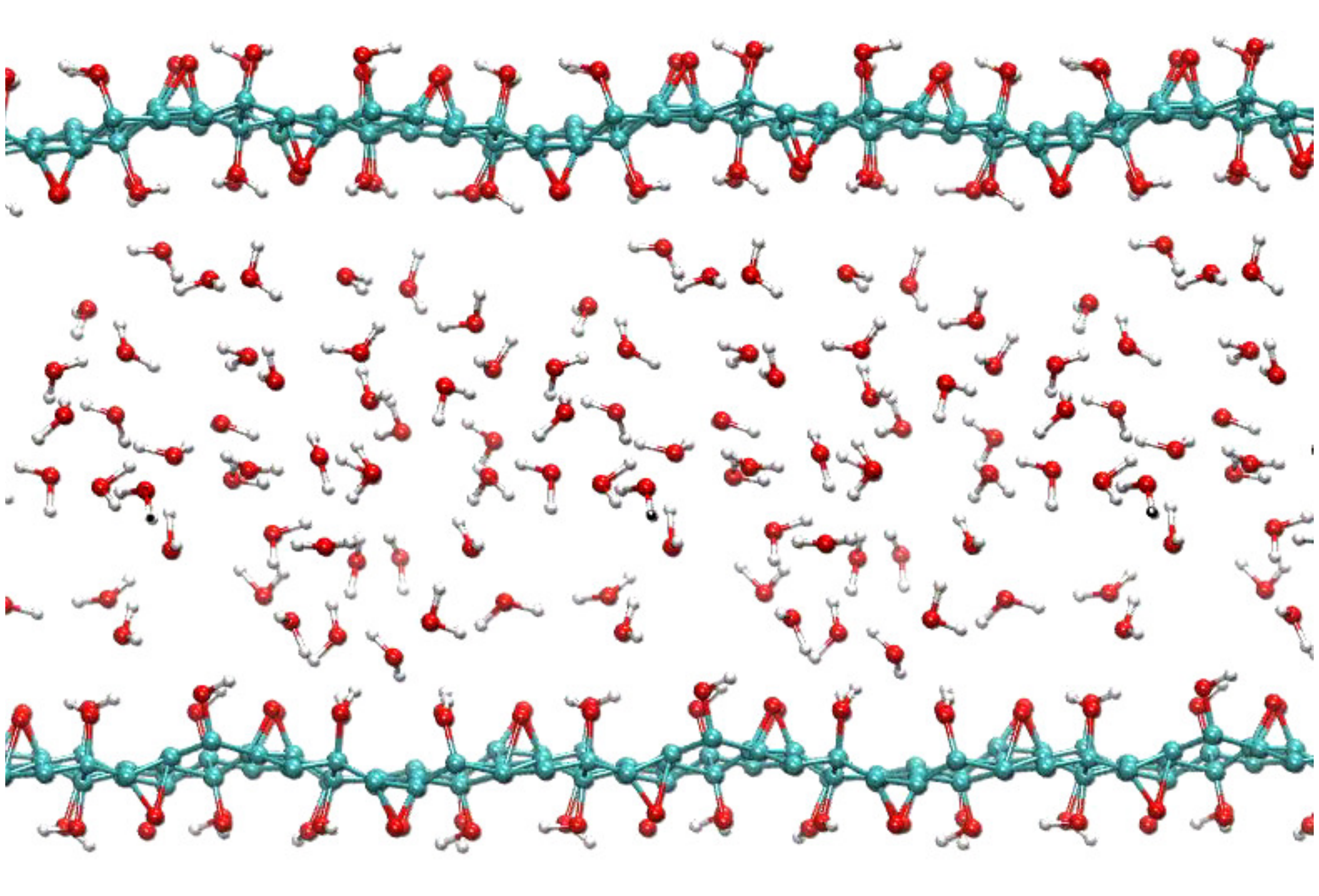}
\setfloatlink{https://nanoten.com/tomanek/manuscripts/DT290v4.mp4} 
\caption{Microcanonical MD simulation of a hydrated GO bilayer
that had been initially equilibrated at $T=1500$~K. %
}
\label{video4}
\end{video}

\begin{video}[h!]
\includegraphics[width=0.55\columnwidth]{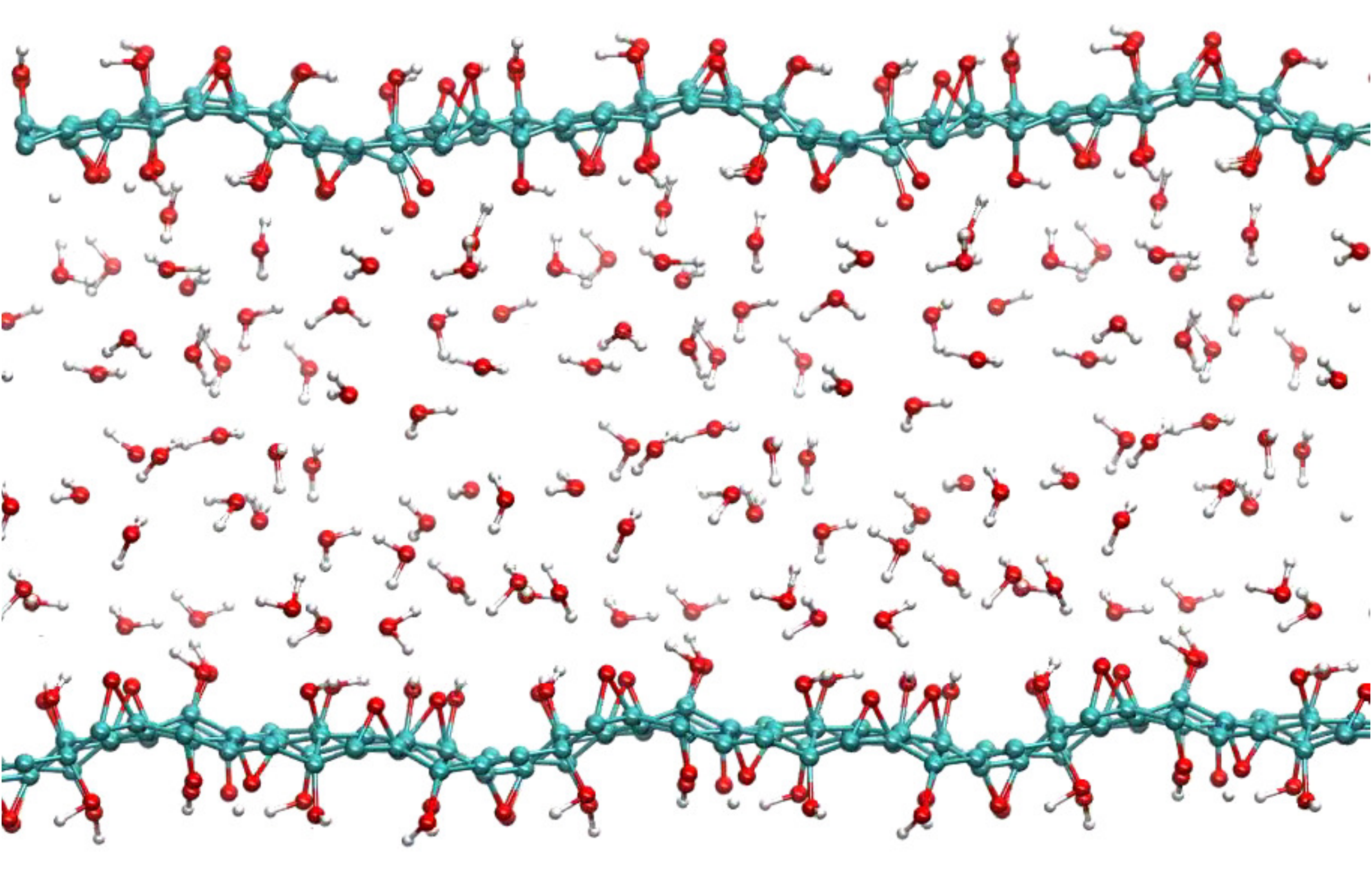}
\setfloatlink{https://nanoten.com/tomanek/manuscripts/DT290v5.mp4} %
\caption{Microcanonical MD simulation of a hydrated GO bilayer
that had been initially equilibrated at $T=2000$~K. %
}
\label{video5}
\end{video}

\begin{video}[h!]
\includegraphics[width=0.60\columnwidth]{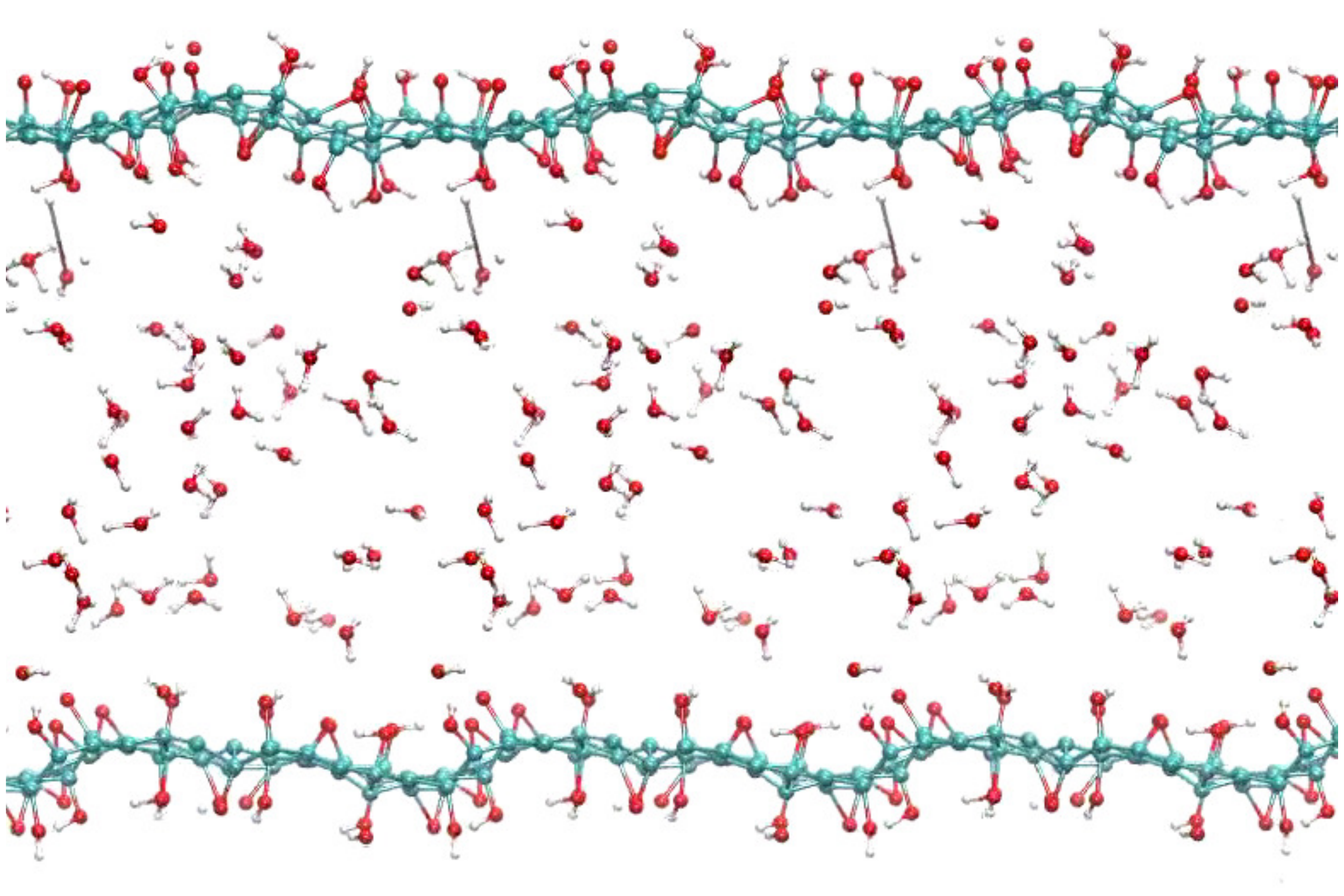}
\setfloatlink{https://nanoten.com/tomanek/manuscripts/DT290v6.mp4} %
\caption{Microcanonical MD simulation of a hydrated GO bilayer
that had been initially equilibrated at $T=2500$~K. %
}
\label{video6}
\end{video}

Results of our MD simulations for the hydrated GO bilayer are
shown in %
Video~\ref{video2} for $T=500$~K, %
Video~\ref{video3} for $T=1000$~K, %
Video~\ref{video4} for $T=1500$~K, %
Video~\ref{video5} for $T=2000$~K, %
Video~\ref{video6} for $T=2500$~K, %
Video~\ref{video7} for $T=3000$~K, %
Video~\ref{video8} for $T=3500$~K, and
Video~\ref{video9} for $T=4000$~K. %
Statistical temperature averages taken during these runs indicate
that the average temperature of a system prepared for %
$T=500$~K has changed to $\left<T\right>=480{\pm}17$~K, %
$1000$~K has changed to $\left<T\right>=1011{\pm}37$~K, %
$1500$~K has changed to $\left<T\right>=1466{\pm}48$~K, %
$2000$~K has changed to $\left<T\right>=2091{\pm}68$~K, %
$2500$~K has changed to $\left<T\right>=2622{\pm}79$~K, %
$3000$~K has changed to $\left<T\right>=2948{\pm}94$~K, %
$3500$~K has changed to $\left<T\right>=3518{\pm}111$~K, and %
$4000$~K has changed to $\left<T\right>=4204{\pm}142$~K %
in our NVE simulations. As stated above, the temperature to
observe a specific process on a natural time scale is
significantly lower than the simulation temperature in our
time-limited study.

\begin{video}[h!]
\includegraphics[width=0.75\columnwidth]{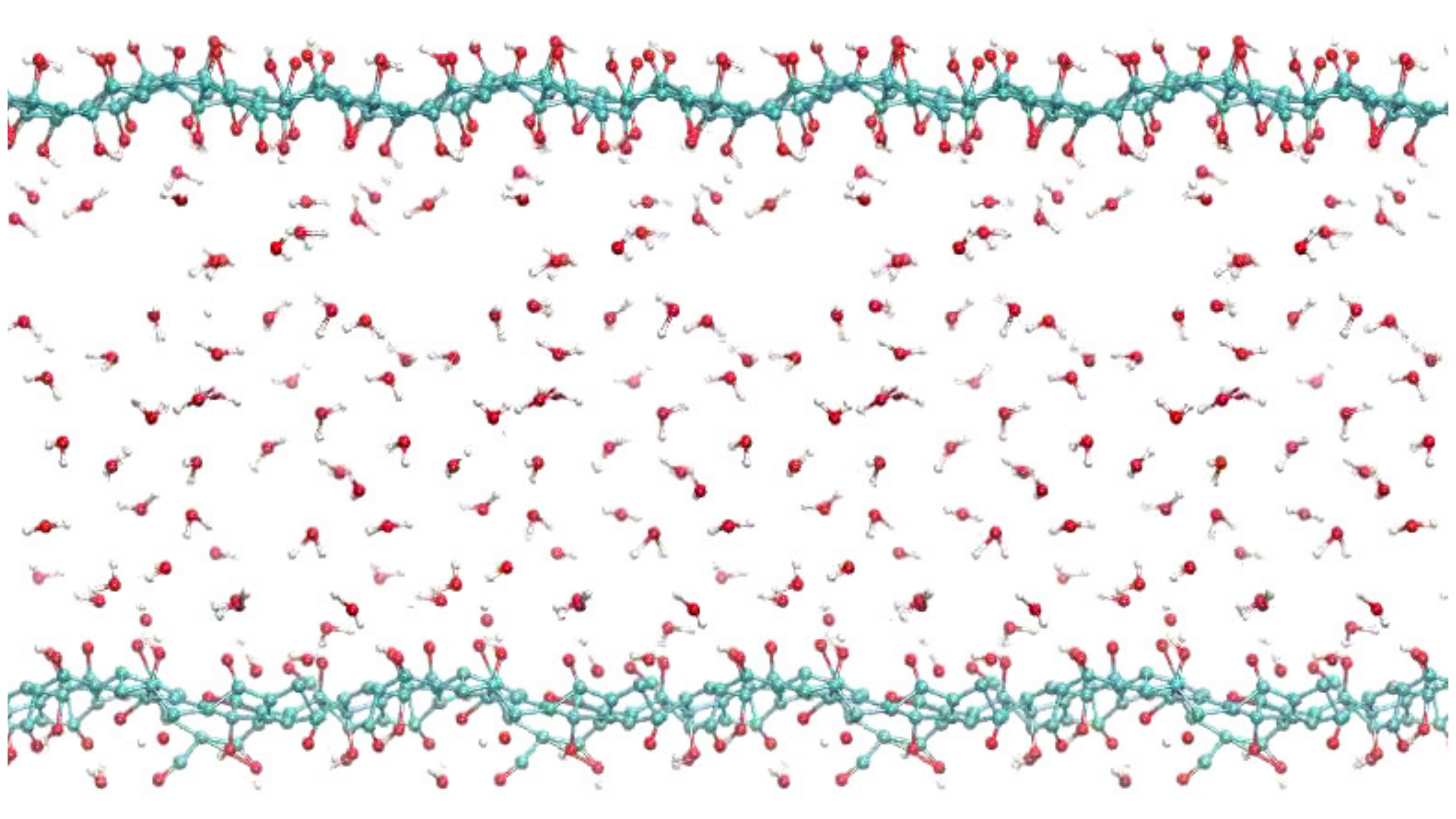}
\setfloatlink{https://nanoten.com/tomanek/manuscripts/DT290v7.mp4} %
\caption{Microcanonical MD simulation of a hydrated GO bilayer
that had been initially equilibrated at $T=3000$~K. %
}
\label{video7}
\end{video}

\begin{video}[h!]
\includegraphics[width=0.80\columnwidth]{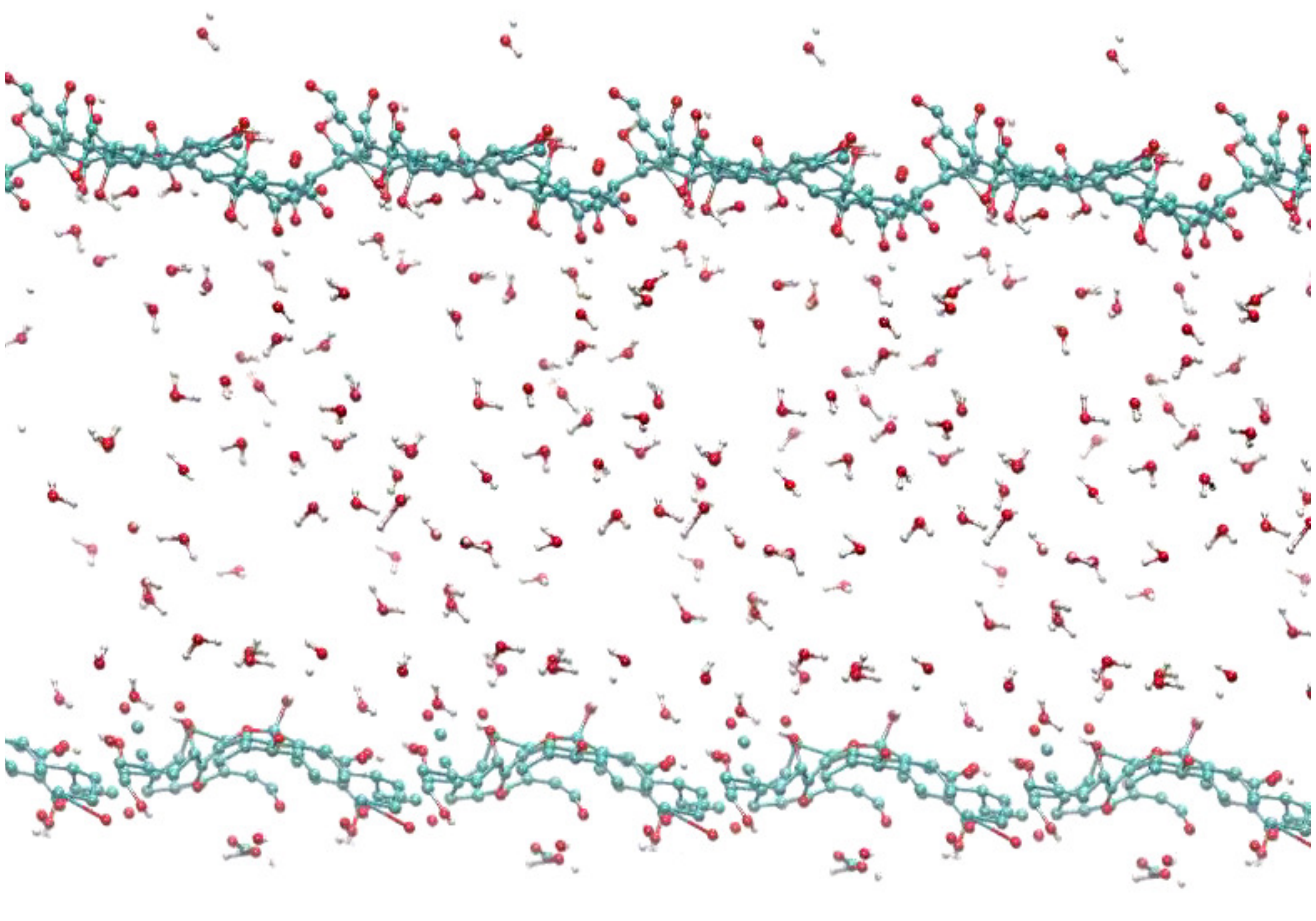}
\setfloatlink{https://nanoten.com/tomanek/manuscripts/DT290v8.mp4} %
\caption{Microcanonical MD simulation of a hydrated GO bilayer
that had been initially equilibrated at $T=3500$~K. %
}
\label{video8}
\end{video}

\begin{video}[h!]
\includegraphics[width=0.80\columnwidth]{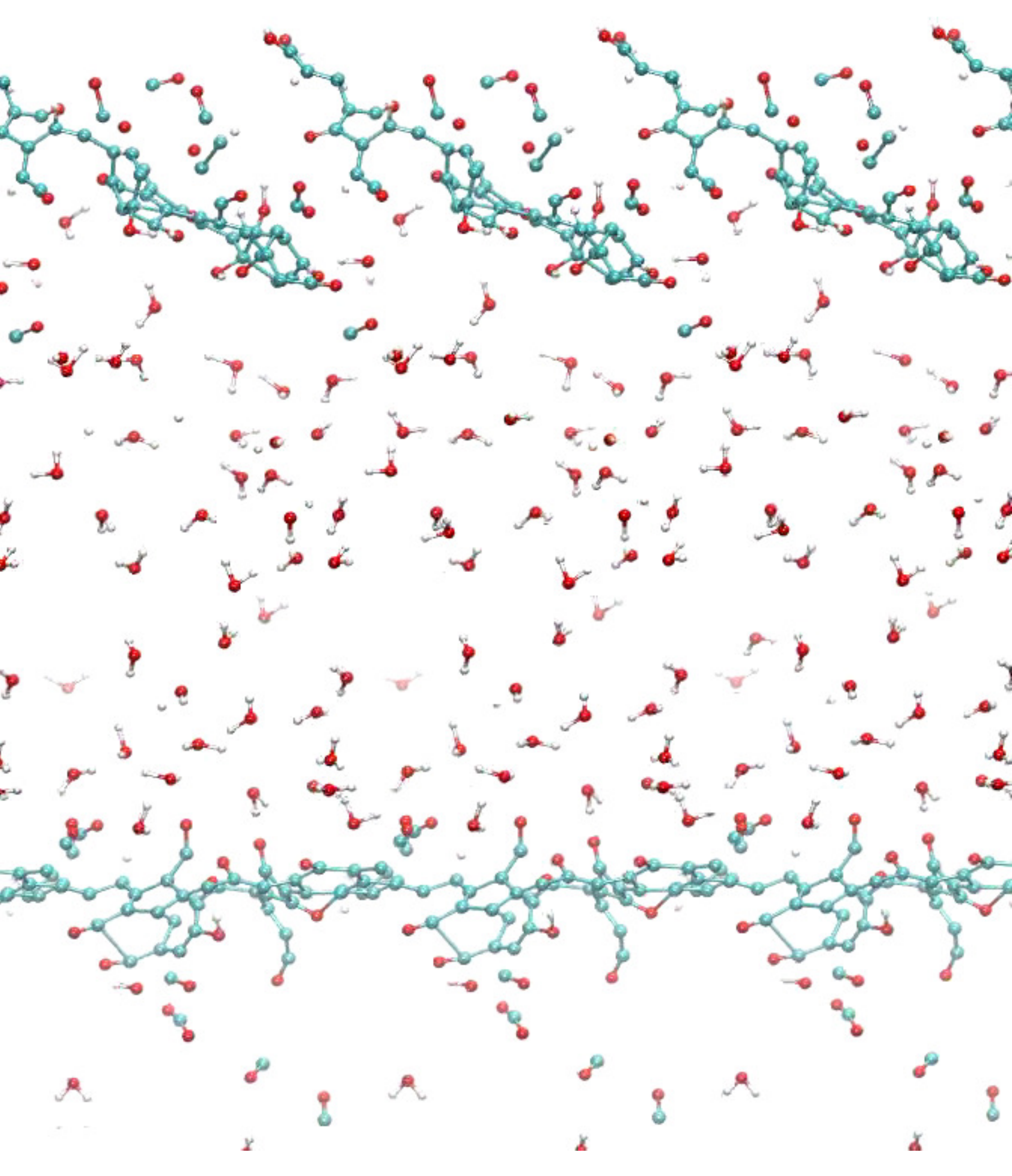}
\setfloatlink{https://nanoten.com/tomanek/manuscripts/DT290v9.mp4} %
\caption{Microcanonical MD simulation of a hydrated GO bilayer
         that had been initially equilibrated at $T=4000$~K. %
}
\label{video9}
\end{video}

These MD simulation results suggest that the GO layers containing
water molecules in the interlayer region remain intact below
$2,500$~K. Nevertheless, these layers are flexible and their
deviation from planarity increases with temperature. With
increasing temperature, liquid water turns to vapor exerting an
increasing pressure on the containing layers and pushing them
apart. At temperatures close to $4,000$~K, the GO layers are
destroyed, whereas the water molecules appear to be unaffected.

Our finding that the graphitic backbone is the most resilient part
of the hydrated GO bilayer is well known, since bare graphene and
graphite are known to to survive temperatures up to
${\approx}3,820$~K~\cite{CRC-CP62}.

\subsection{Swelling at elevated temperatures}

The effect of increasing water pressure on the interlayer distance
$d_{il}$ at elevated temperatures is shown in Fig.~\ref{fig3}. For
a given simulation temperature $T$, we plot the time evolution of
the average interlayer distance $\left<d_{il}\right>(t)$, defined
as an average separation $d$ between C atoms in the top and bottom
GO layer in the direction normal to the plane of the bilayer. The
error bars indicate the width of the distribution of $d$ values
across the unit cell. At a constant water pressure, which
increases with temperature, we expect a constant acceleration,
resulting in a parabolic dependence of the inter-layer distance
$d_{il}$ on time $t$ if we ignore the pressure drop during the
short time period considered. Quadratic fits to
$\left<d_{il}\right>(t)$ are indicated by the dotted lines in
Fig.~\ref{fig3}. As expected, the acceleration of the interlayer
separation, reflected in the harmonic coefficients, increases with
increasing temperature due to the increasing water vapor pressure.

\begin{figure}[t]
    \centering
    \includegraphics[width=1.0\columnwidth]{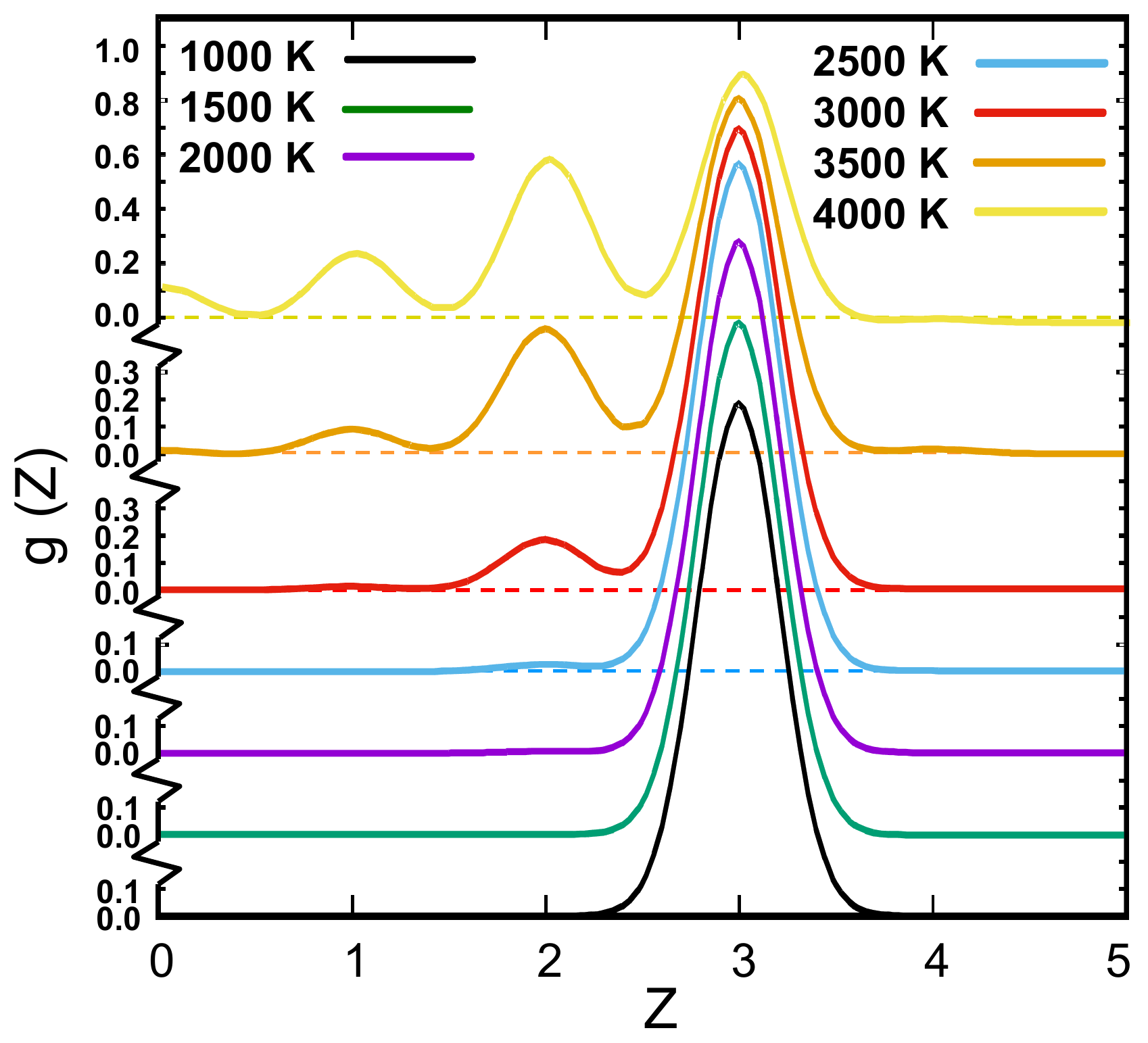}
    \caption{%
        Average distribution of coordination numbers $Z$ within
        the C backbone of the bilayer as a function of
        the simulation temperature.
        For better comparison, the discrete distribution has been
        convoluted by a Gaussian with a FWHM=0.5.}%
    \label{fig4}
\end{figure}

\subsection{Coordination numbers at elevated temperatures}

The simplest way to investigate the intactness of the graphitic
backbone is a simple study of the coordination number $Z$
distribution within the layer. In our study, we went through all
96 C atoms in the unit cell and counted all other C atoms that
were closer than $1.94$~{\AA}, the average between the nearest and
the second nearest distance in graphene, as nearest neighbors.
This gives a discrete histogram for every simulation. Instead of
an awkward comparison between 7 histograms, we convoluted the
$\delta-$functions of different strength at integer values of $Z$
by a Gaussian of FWHM=$0.5$ and present the corresponding
distribution $g(Z)$, with $Z$ a continuous variable now, in
Fig.~\ref{fig4} for the different temperatures. Since the integral
over $g(Z)$ is normalized to 1, the area of a peak at $Z_i$
indicates the probability for any C atom to have the coordination
number $Z_i$.

Our results show a single peak at $Z=3$ for temperatures
$T{\alt}2,000$~K, indicating that all carbon atoms maintain a
graphene-like local environment with three nearest neighbors. At
$2,500$~K we observe the emergence of a new peak at $Z=2$, which
becomes more prominent at $3,000$~K, indicating that some C-C
bonds have been broken. $Z=2$ carbon atoms are found at the edge
of graphitic flakes or within linear chains. A new peak at $Z=1$
emerges at $T{\agt}3,500$~K, indicating C atoms at the end of a C
chain. Finally, at $T{\agt}3,500$~K, we observe the emergence of
isolated C atoms with $Z=0$ that are disconnected from the
backbone.

\begin{figure}[h]
    \centering
        \includegraphics[width=1.0\columnwidth]{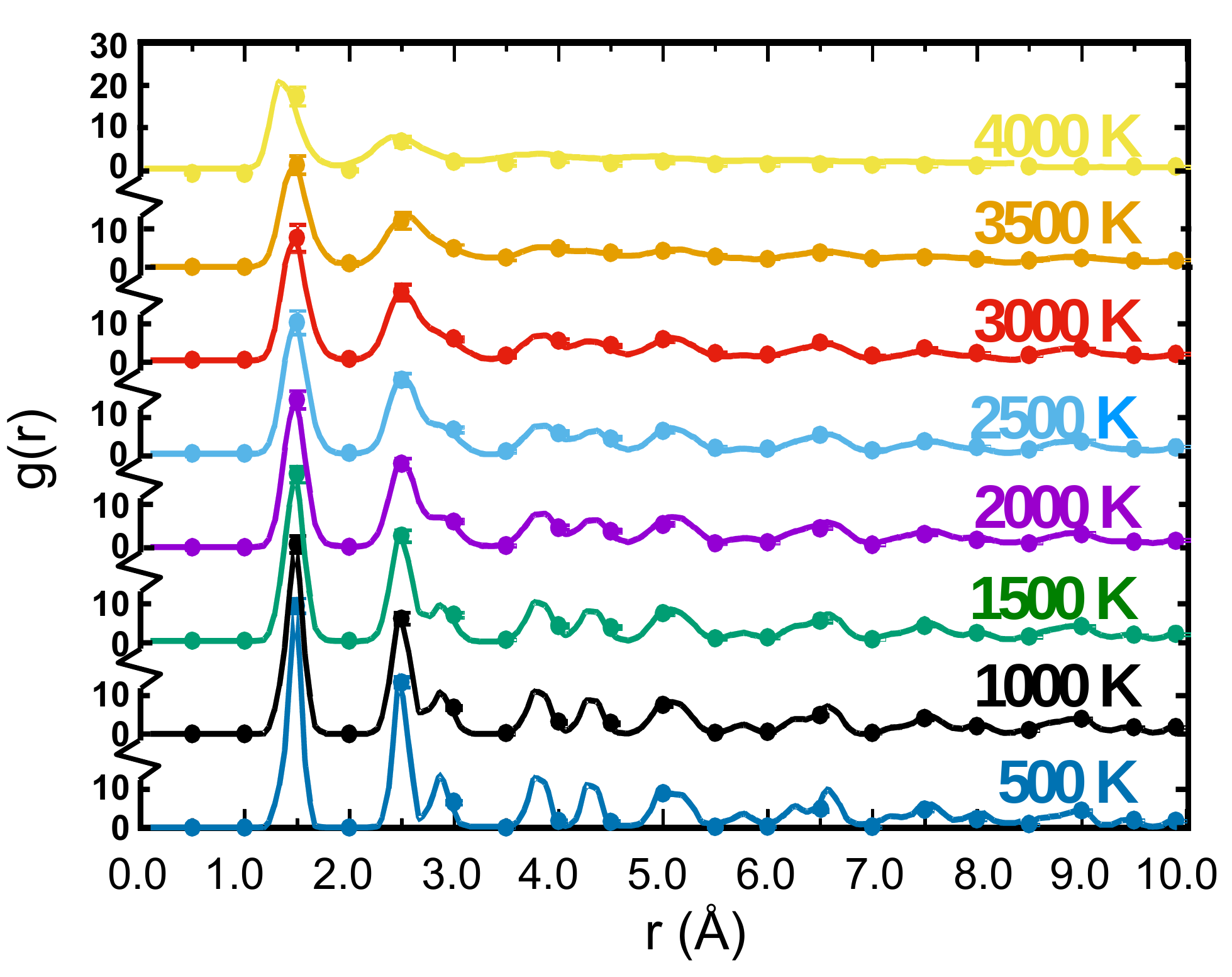}
    \caption{%
        Average C-C pair correlation function
        $\left<g(r)\right>$ of the system as a function of
        the initial simulation temperature $T$,
        convoluted with a Gaussian
        with a FWHM=$0.1$~{\AA}.
        }%
    \label{fig5}
\end{figure}

\subsection{Pair correlation function at elevated temperatures}

A well-defined quantity to characterize the structure of the
graphitic backbone is the pair correlation function $g_{C-C}(r)$,
which indicates the probability to find a C neighbor within a thin
spherical shell of radius $r$ around a C atom. Unlike the
above-defined $g(Z)$, $g_{C-C}(r)$ can be observed by X-ray or
e-beam diffraction. In a perfect sheet of graphene frozen at
$T=0$, $g_{C-C}(r)$ consists of a series of $\delta-$functions,
which get broadened and modified at higher temperatures. We
display the time averaged pair correlation function
$\left<g_{C-C}(r)\right>$ as a function of the distance $r$,
obtained during our simulations, in Fig.~\ref{fig5} for MD runs at
different simulation temperatures. Our results indicate that the
shape of $g_{C-C}(r)$ does not change much from that of graphene
for $T{\alt}2,000$~K. At higher temperatures, however,
$\left<g_{C-C}(r)\right>$ smoothes significantly, especially at
larger interatomic distances.

\begin{figure}[b]
    \centering
    \includegraphics[width=1.0\columnwidth]{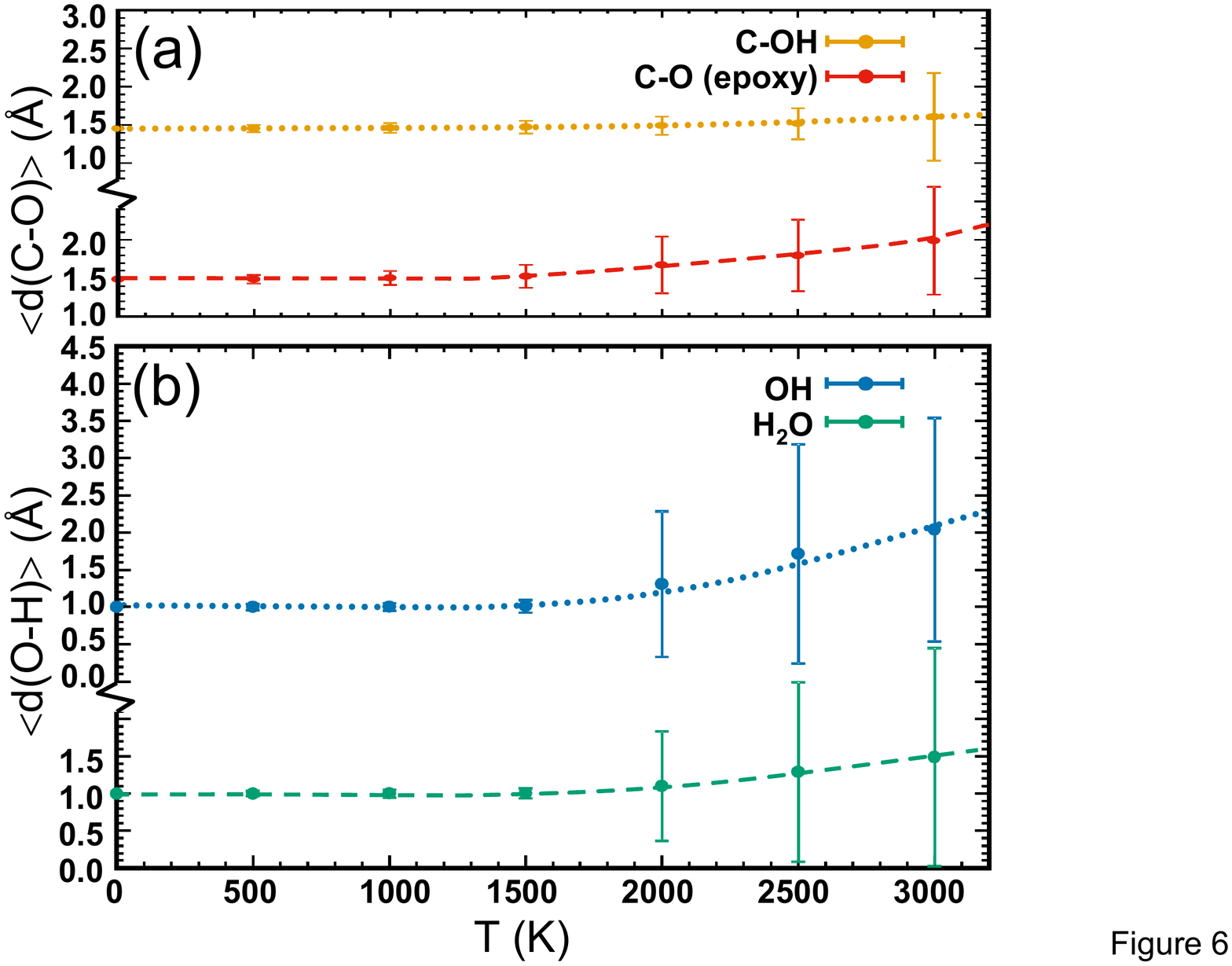}
    \caption{%
        (a) Average C-O distance $\left<d(C-O)\right>$ between
        oxygen atoms forming OH groups (dotted line) or
        epoxy bonds (dashed line) and their closest carbon
        neighbors as a function of the initial simulation
        temperature $T$.
        (b) Average O-H distance $\left<d(O-H)\right>$ between
        oxygen atoms forming OH groups (dotted line) or
        water molecules (dashed line) and their closest hydrogen
        neighbors as a function of $T$.
        Large error bars in the distribution indicate detachment
        or exchange of atoms with nearby molecules or radicals. }%
    \label{fig6}
\end{figure}

\subsection{Changes in functional groups and H$_2$O at
            elevated temperatures}

One of the important questions, to which our study provides an
answer, is whether the epoxy-O and OH functional groups remain
attached or whether they detach from the graphitic backbone before
it thermally disintegrates. This information is important, since
only functional groups that are chemically bonded to the substrate
will weaken its structure locally and lower its melting
temperature. A useful guideline for the hierarchy of thermally
activated processes is
provided by comparing the relevant bond energies. These amount to %
$3.1-4.0$~eV for a single C-C bond, %
$4.1$~eV for a C-O bond, and %
$4.5$~eV for an O-H bond~\cite{H2O-CRC86}.
Our results for the spread of interatomic distances in
Fig.~\ref{fig6} are indeed consistent with the hierarchy of bond
strengths and indicate that the functional groups generally stay
chemisorbed to a defect-free graphitic backbone until its thermal
destruction.

Exchange of hydrogen atoms between water molecules occurs even at
$0$\textdegree{C}, as evidenced by investigations of
D$_{2}$O/H$_{2}$O ice interfaces~\cite{Drori2017}. Earlier
SCC-DFTB studies of ``dry'' GO have indicated that hydrogen
migrations occurs in this material at ambient
temperatures~\cite{Paci2007}. In our simulations we have observed
the first hydrogen hopping events at $2,500$~K, which represent
both OH-water and OH-O-epoxy exchanges. As mentioned above, the
high simulation temperature is necessitated by the limited
duration of our thermally accelerated MD study. With no time
limit, the same processes will occur at much lower temperatures in
nature.

Oxygen-carbon bonds survive up to even higher temperatures
\textendash OH-groups remain attached to the graphitic layers up
to $3,000$~K, followed by the detachment of epoxy-O groups at
$3,500$~K. Some migration of oxygen atoms was observed already at
$2,500$~K and will be discussed in the Appendix. SCC-DFTB
simulations have reported oxygen atom migration at
$1,323$~K~\cite{Paci2007}, which is close to the exfoliation
temperature of GO in theoretical studies. Some results indicate
agglomeration of oxygen-containing groups to form highly oxidized
areas surrounded by nanodomains of pristine
graphene~\cite{Zhou2014} during GO aging. The small unit cell size
and limited simulation time period do not allow us to judge if
this process really occurs.

Epoxide groups were earlier shown to be of paramount importance
for the thermal decomposition of GO sheets, since C-C bond in an
epoxy-group breaks easier than a regular graphitic bond. According
to DFT calculations, barriers for this process are mostly less
then $1.0$~eV~\cite{Sun2012}. Aggregation of epoxy-groups lower
the barrier to ${\approx}0.6$~eV. In case of a linear arrangement
of such defects, a crack in the carbon sheet is formed in a
process dubbed O-driven ``unzipping''~\cite{Li06}. We observed C-C
bond breaking in epoxy-groups already at $3,000$~K in our
simulations.

Our simulations indicate a total destruction of the GO backbone at
$T{\approx}4,000$~K, accompanied by the detachment of small
molecules as by-products. These by-products are dominated by
carbon monoxide, in-line with previously published
results~\cite{Paci2007}, as well as carbon dioxide and water. As
during the thermal disintegration of other carbon
allotropes~\cite{DT080}, free-standing chains of several carbon
atoms are formed, stabilized by their high entropy.


\section{Discussion}

As mentioned earlier, our attempt to provide atomic-scale insight
into processes occurring in hydrated GO at elevated temperatures
poses fundamental challenges and must be seen only as a first step
towards obtaining microscopic understanding. The first limiting
factor is the observation time. Activated processes including
oxidation and reduction of solid surfaces occur on the time scale
of seconds, whereas state-of-the art \textit{ab initio}
calculations for unit cells containing few hundred atoms may
address only ${\alt}10^{-12}-$second time intervals, requiring --
as in our study -- several months of CPU time on massively
parallel supercomputers. We tried to address this problem using
thermally accelerated dynamics, namely by exposing the system to
an artificially increased external temperature. Even though this
approach does not allow us to estimate, at which temperature a
particular process would occur in nature, it allows us to judge if
specific reactions, such as migration of functional groups and
their eventual detachment from the GO substrate, occur at lower or
similar temperatures as the disintegration of the GO backbone that
involves breaking of C-C bonds.

It is true that the \textit{ab initio} approach is time consuming.
For this reason, the vast majority of atomistic MD simulations is
based on parameterized force fields, which offer a higher degree
of numerical efficiency and allow to study the motion of several
thousand atoms simultaneously. The CPU speed gain may be up to
three orders of magnitude, but still falls short of the twelve
orders of magnitude needed to observe relevant reactions. The
drawback of parameterized force fields is their lack of
universality and quantitative predictability, which may lead to
incorrect conclusions. In particular, force fields optimized for
bulk fluids need to be changed at interfaces and in situations,
where long-range electrostatic interactions play a nontrivial
role~\cite{{Kohler2019},{Misra2017a}}. In spite of its high
computational demand, DFT is nominally free of parameters and
independent of predefined assumptions. The approach used in this
study has been validated, among others, by successfully predicting
static and dynamic properties of liquid water~\cite{DT274}, and
should provide valuable information that should complement
large-scale studies with parameterized force fields.

The second factor limiting our study is the arrangement of small
graphitic flakes, functionalized by epoxy and hydroxy groups and
separated by water, in realistic GO material that is lacking
long-range order. Our study focusses on processes occurring on a
pair of adjacent GO flakes separated by water, but ignores their
finite size. Samples with different arrangements of GO flakes and
different types of defects may behave very differently at high
temperatures.

Even though the presence of water in the interlayer region does
not affect the dynamics of the GO layers, H$_2$O molecules should
in no way be considered as inert. Selected reactions we discuss in
the Appendix indicate that hydrogen exchange between water
molecules and epoxy or hydroxy functional groups are rather common
at elevated temperatures, still leaving water molecules among the
stable products.

After revealing the strong points and limitations of our
theoretical study, we must summarize what we have really learned.
Many of our findings confirm conclusions based on chemical
intuition. Water molecules remain intact up to the disintegration
point of the GO backbone, initiated by C-C bond breaking near
adsorbed functional groups. At high enough temperatures, liquid
water turns into vapor that exerts a significant pressure of GO
layers, separating them in a process known as
swelling~\cite{{Talyzin2008},{Talyzin2011},{Talyzin21-swell}}.
Presence of oxygen in functional groups is the microscopic reason
for GO -- unlike graphene and graphite -- being hydrophilic. We
find this to be the case in the Lerf-Klinowski
model~\cite{Lerf1998} of GO underlying our study, where the ratio
of carbon to epoxy-O is 6. Functional groups may move across the
GO substrate, often by intricate reactions involving adjacent
water molecules. Still, epoxy and hydroxy groups do not desorb
below the disintegration temperature of the GO backbone. At that
point, these functional groups migrate to the graphitic edges
exposed after fracture and terminate them.

We went to great lengths to automatically extract useful
information from the vast number of atomic coordinate files
collected during our MD simulations. Information obtained in this
way includes statistical data about temperature fluctuations in
Fig.~\ref{fig2}, interlayer distances in Fig.~\ref{fig3}, atomic
coordination numbers in Fig.~\ref{fig4}, the C-C pair correlation
function in Fig.~\ref{fig5}, and separation of functional groups
from the substrate in Fig.~\ref{fig6}. Our collected data for
coordination numbers, pair correlation functions as well as C-O
and O-H distances clearly show that the structure of hydrated GO
remains essentially unchanged up to the simulation temperature
$T{\approx}2,000$~K. The only notable processes occurring below
this temperature are GO swelling driven by water pressure, and
buckling of the GO layers. Both of them show increasing amplitudes
with increasing temperature.

We then subjected our collected structural information to a data
mining analysis to search for unusual processes. We found that
such unusual processes, which deserve the attribute of ``chemical
reactions'',  start occurring only at $T{\approx}2,000$~K. These
include the C-C bond fracture leading to under-coordinated carbon
atoms with the coordination number $Z<3$ and hydrogen exchange
reactions, traced by the $\left<d(C-O)\right>$ distribution, which
occur more frequently at high temperatures. Structural
deterioration of the carbon substrate becomes particularly visible
in the smoothing of the C-C pair correlation function
$\left<g(r)\right>$ at high temperatures.

Starting at $T{\approx}2,500$~K, we observe various reactions
involving hydroxyls and epoxy-groups chemisorbed on GO layers and
water molecules in the interlayer region. Reactions of interest,
many of which are detailed in the Appendix, include hydrogen
exchanges, inter-conversion between OH- and epoxy-groups,
formation of peroxides, 1,3-epoxides and monocoordinated oxygens.

Our study of a defect-free model system, a hydrated GO bilayer,
does not address the reorientation of GO flakes in presence of
water that is free to escape through in-layer defects. Thus,
dehydration and reduction of GO to rGO at high temperatures is
beyond the scope of this study. Specific arrangements of GO flakes
may block off compartments of different size within the system.
Water contained in such compartments will exert pressure and
eventually burst the containing structure at high enough
temperatures. This process, known as deflagration, may occur
across a wide temperature range, as also evidenced in the
experiment~\cite{Boehm1965}.

Our thermally accelerated dynamics simulations, while revealing
commonly occurring reactions, do not allow us to estimate the
temperature, when such reactions occur in nature on an unlimited
time scale. As discussed earlier, the nominal simulation
temperature may be seen as the upper limit of the expected
temperature range, with a realistic value being
${\approx}200-2,000$~K lower. Still, we expect the sequence of
thermal onsets of specific reactions to remain the same in our
$1$-ps simulations and, at reduced temperatures, under
experimental conditions on a much longer time scale. For realistic
temperature estimates, we may take recourse to what is known from
the experiment.

High-temperature behavior of GO, specifically its thermal
reduction, has been studied experimentally in a vast number of
publications, with most attention devoted to Hummers GO. A
comparative study of many GO samples produced in different ways
revealed significant differences in their
behavior~\cite{{Boehm1965},{Feicht2017}}. Very impressive
variations occur in the deflagration temperatures ranging from
$194-325$\textdegree{C}, with Hummers GO covering the lower end
and Brodie GO the upper end of the temperature range. Most
differences in behavior are caused by the size and
functionalization of GO flakes on their faces and edges, flake
arrangements within the sample, and presence of water. Thermal
decomposition of GO is an exothermic process, which is slowed down
below the deflagration point by the evaporation of
water~\cite{Boehm1965}.

Experimental data indicate that water escapes from GO above
$100$\textdegree{C}~\cite{Talyzin-PC}, followed by exfoliation in
the temperature range between
$200-300$\textdegree{C}~\cite{{You2013},{Boehm1965}}. Along with
loss of water~\cite{Boehm1965}, almost all oxygen-containing
groups have been shown to disappear after heating to temperatures
above $200$\textdegree{C}~\cite{Hu2012}. At these temperatures, GO
has transformed to hydrophobic rGO, essentially graphitic carbon,
with no swelling propensity~\cite{Talyzin-PC}.

Comparison between processes observed at a particular temperature
and our thermally accelerated simulation results indicates that
the simulation temperature of $T=2,500$~K roughly corresponds to
the temperature range of $100-200$\textdegree{C} that precedes
formation of rGO.

At still higher temperatures, most interesting processes involve
changes in the functional groups within rGO. Oxygen groups
decorating the faces are likely to migrate towards the reactive
edges of disintegrating GO, where they survive as carboxyl and
carbonyl groups up to
$500-700$\textdegree{C}~\cite{{Talyzin-PC},{Deemer17}}. This
finding is consistent with results of thermogravimetric analyses
of GO samples, which were prepared by various
methods~\cite{Deemer17} and displayed no weight loss above
$600$\textdegree{C}~\cite{{Hu2012}, {Farivar2021}}. Analysis of
electron energy-loss spectroscopy (EELS) and X-ray photoelectron
spectroscopy (XPS) data of GO heated to different temperatures has
revealed that epoxide groups disappear above $400$\textdegree{C},
whereas hydroxyl groups remain present up to
$1,000$\textdegree{C}~\cite{DAngelo2017}.

Further temperature increase leads to the detachment of other
functional groups and C-C bond fracture that initiates a gradual
disintegration of GO. These processes occur at the simulation
temperature of $4,000$~K, which should roughly correspond to
$T{\alt}1,000$~K under realistic experimental conditions with no
time constraints. Since processes at this stage are irreversible,
this temperature range must be avoided during a thermal treatment
of GO membranes.

Even though many experimental results related to the thermal
decomposition of GO are controversial, there is general consensus
about the major trends. Differences between published studies are
likely related to inherent variations in the microstructure of GO
samples produced by different techniques and subjected to
different heat treatment protocols~\cite{Deemer17}.
\\


\section{Summary and Conclusions}

In summary, we have performed {\em ab initio} density functional
theory molecular dynamics simulations addressing thermally driven
structural changes in a bilayer of hydrated graphite oxide in
vacuum. This rather artificial system was selected as a model to
study microscopic details of the swelling and thermal
decomposition of hydrated graphite oxide (GO), which has been
demonstrated to allow water permeation while rejecting solvated
ions in the feed~\cite{Boehm1961} when sandwiched in-between
layers of carbon nanotube buckypaper and carbon fabric for
containment and mechanical strength~\cite{DT274}. In order to
observe slow processes in the short time frame of the simulation,
we artificially raised the system temperature in our thermally
accelerated MD studies of a perfect GO bilayer. Covering the
temperature range~\cite{MD-temperature} up to $4,000$~K, we find
that water molecules in the interlayer region are rather decoupled
from the GO layers and only marginally affect their behavior at
moderate temperatures, except for increasing swelling by water
vapor pressure. In presence of nearby water molecules, some
epoxy-O atoms move from their bridge to the on-top site, turning
into radicals and changing the configuration of the connected
carbon atom from $sp^3$ to $sp^2$. In a similar way, in the
vicinity of water, hydrogen atoms may detach from adsorbed OH
groups, converting them to epoxy-groups. Both processes facilitate
buckling and local fracture of the graphitic backbone above
$2,500$~K. At higher temperatures, we observe the destruction of
the graphitic backbone itself. Oxygens in the functional groups
migrate from the faces to the reactive exposed edges of the
graphitic flakes. Depopulation of oxygen on the face of graphitic
flakes turns GO into hydrophobic reduced GO (rGO) not subject to
swelling.

Combining knowledge contained in a vast number of experimental
studies with the microscopic insight provided by our simulations,
we find GO not only very promising for water treatment and
desalination~\cite{{Boehm1961},{DT274}}, but also thermally very
stable. Thus, exposing GO membranes -- in an inert gas atmosphere
containing water vapor -- to temperatures below
$300$\textdegree{C} or even higher may offer a viable alternative
to chemical cleaning for removing biofouling residue.


\renewcommand\thesubsection{\Alph{subsection}}
\renewcommand{\theequation}{A\arabic{equation}}
\renewcommand{\thevideo}{A\arabic{video}}
\setcounter{subsection}{0} %
\setcounter{equation}{0} %


\section*{Appendix}

\subsection{Atomic-scale reactions at the surface of hydrated GO}

\begin{video}[h]
\includegraphics[width=0.30\columnwidth]{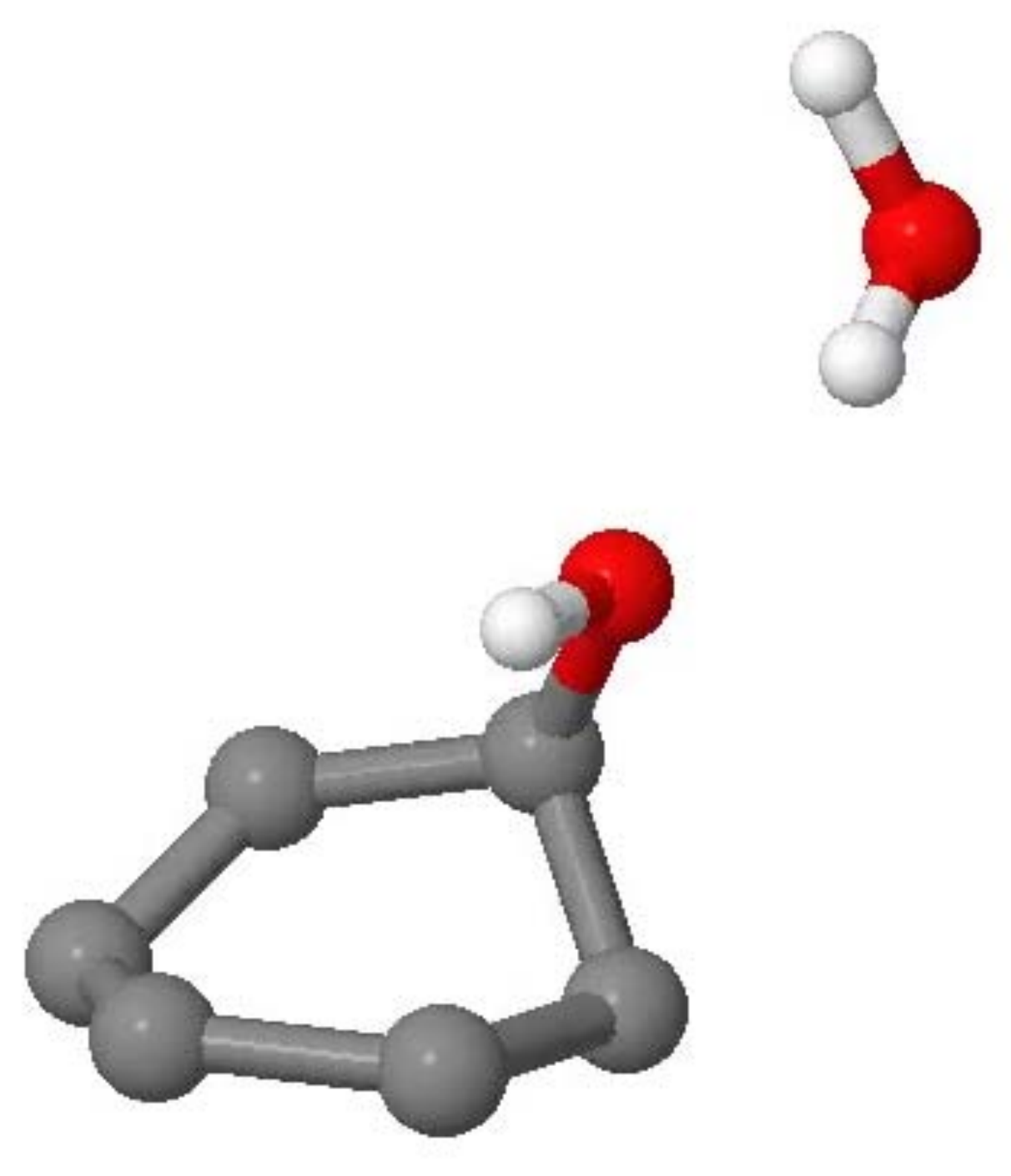}
\setfloatlink{https://nanoten.com/tomanek/manuscripts/DT290vA10.mp4} %
\caption{Interaction of a water molecule with an OH-group attached
to GO at the simulation temperature $2,000$~K.%
}
\label{videoA10}
\end{video}

We have subjected the trajectory data of our MD simulations to an
automated data mining process in order to learn about interesting
reactions that take place. These reactions are discussed in the
following.

Video~\ref{videoA10} demonstrates the attraction of a water
molecule to a hydroxyl group attached to GO, one of the reasons
that turn GO hydrophilic, at 2,000~K. The dynamics of the rather
trivial reaction in this video, which depicts temporary attachment
of H$_2$O to the oxygen in the hydroxyl group, illustrates the
role of hydrogen bonds, which are essential for the structure of
ice and liquid water. These bonds constantly emerge and break in
liquid water.

\begin{video}[h]
\includegraphics[width=0.30\columnwidth]{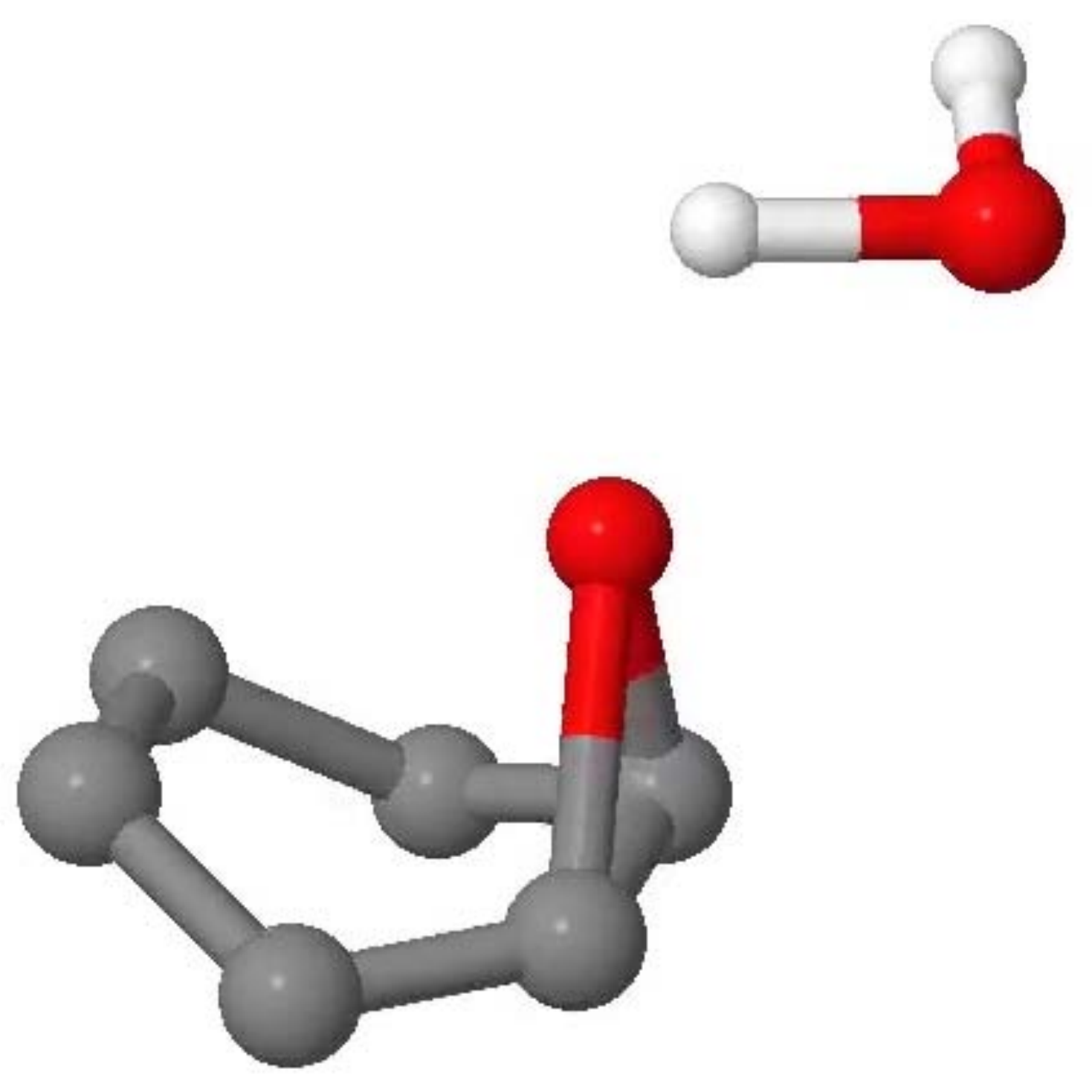}
\setfloatlink{https://nanoten.com/tomanek/manuscripts/DT290vA11.mp4} %
\caption{Interaction of a water molecule with an epoxy-group
attached to GO at the simulation temperature $2,500$~K.%
}
\label{videoA11}
\end{video}

Video~\ref{videoA11} depicts a very different process, namely C-O
bond splitting in an epoxide group at 2,500~K, followed by a
hydrogen transfer from a nearby H$_2$O molecule. The hydrogen
transfer is expected based on experimental data, which suggest
that alkyl-OH groups are stronger acceptors of hydrogen bonds than
water while being comparable donors~\cite{Abraham1989}. Thus, we
may expect more frequent formation of long-lasting H-bonds with
water being the donor and OH-group being the acceptor, than the
other way around. Epoxy-groups, which have alkyl substituents,
should then serve as even better hydrogen bond acceptors. The
process starts with a vibration in the three-membered ring
constituting the epoxy-group, during which one of the
carbon-oxygen bonds breaks, releasing an estimated strain energy
of 0.28~eV~\cite{Whalen2005}. The single-coordinated oxygen atom
then captures a hydrogen atom from an H$_2$O molecule. As a
result, the initial epoxy-group is converted to an OH-group,
whereas a water molecule turns into an OH anion.

\begin{video}[h]
\includegraphics[width=0.20\columnwidth]{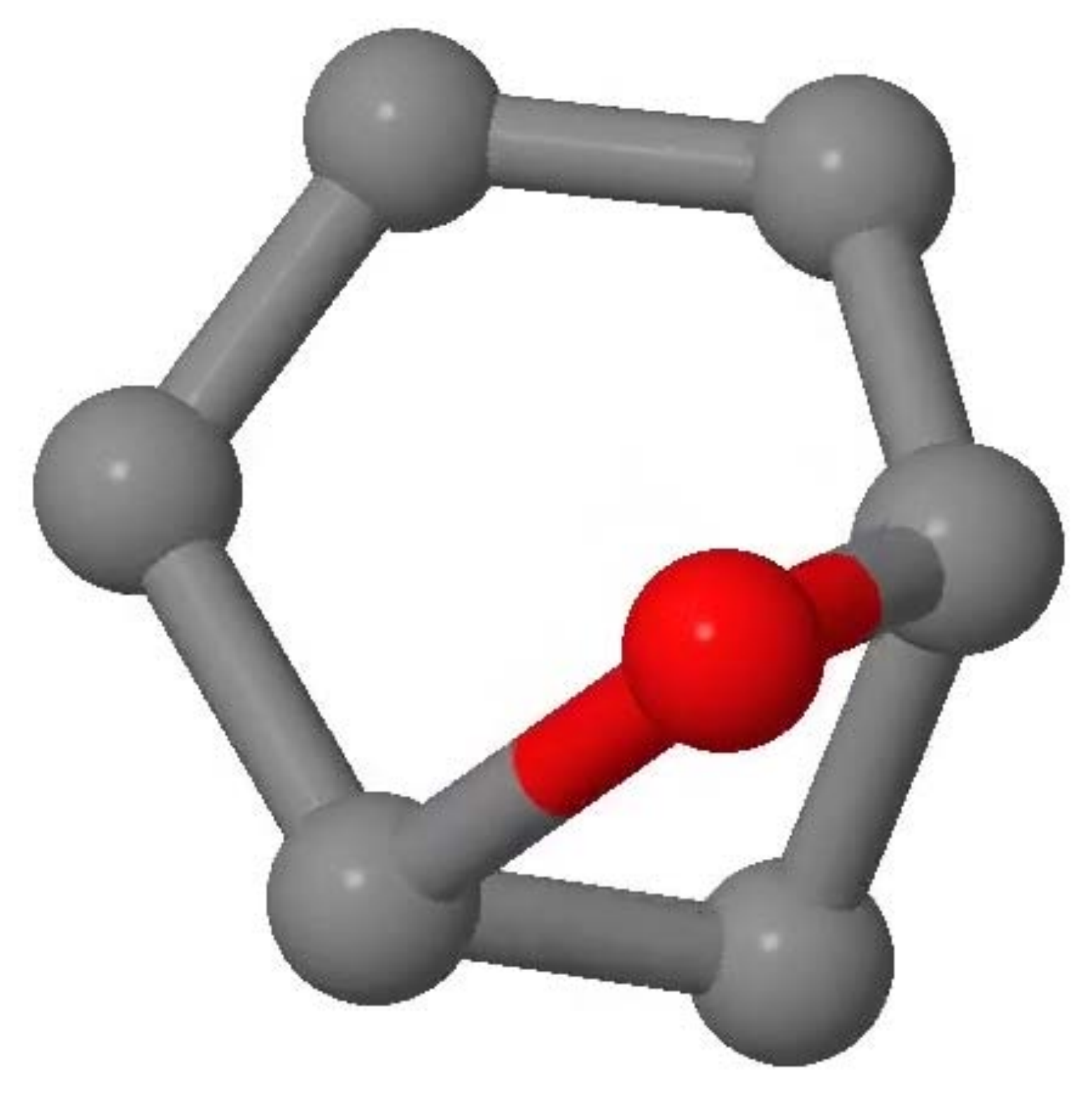}
\setfloatlink{https://nanoten.com/tomanek/manuscripts/DT290vA12.mp4} %
\caption{Formation of a 1,3-epoxide group and substrate
destabilization
at the simulation temperature $2,500$~K.%
}
\label{videoA12}
\end{video}

Video~\ref{videoA12} illustrates unusual epoxy-bond scenarios
possible on a flexible substrate. During substrate vibrations,
second neighbors in the graphene lattice may come close enough to
bind with a ``dangling'' oxygen and thus form an 1,3-epoxide
group. This structure appears to be rather stable at 2,500~K, as
also evidenced by its occurrence in nature, namely as a part of
cytotoxic triterpenes sodwanones I and W from marine sponge
species~\cite{Dai2006, Rudi1995}. Eventually, the strain energy is
released by breaking a C-C bond in the substrate.

\begin{video}[h]
\includegraphics[width=0.30\columnwidth]{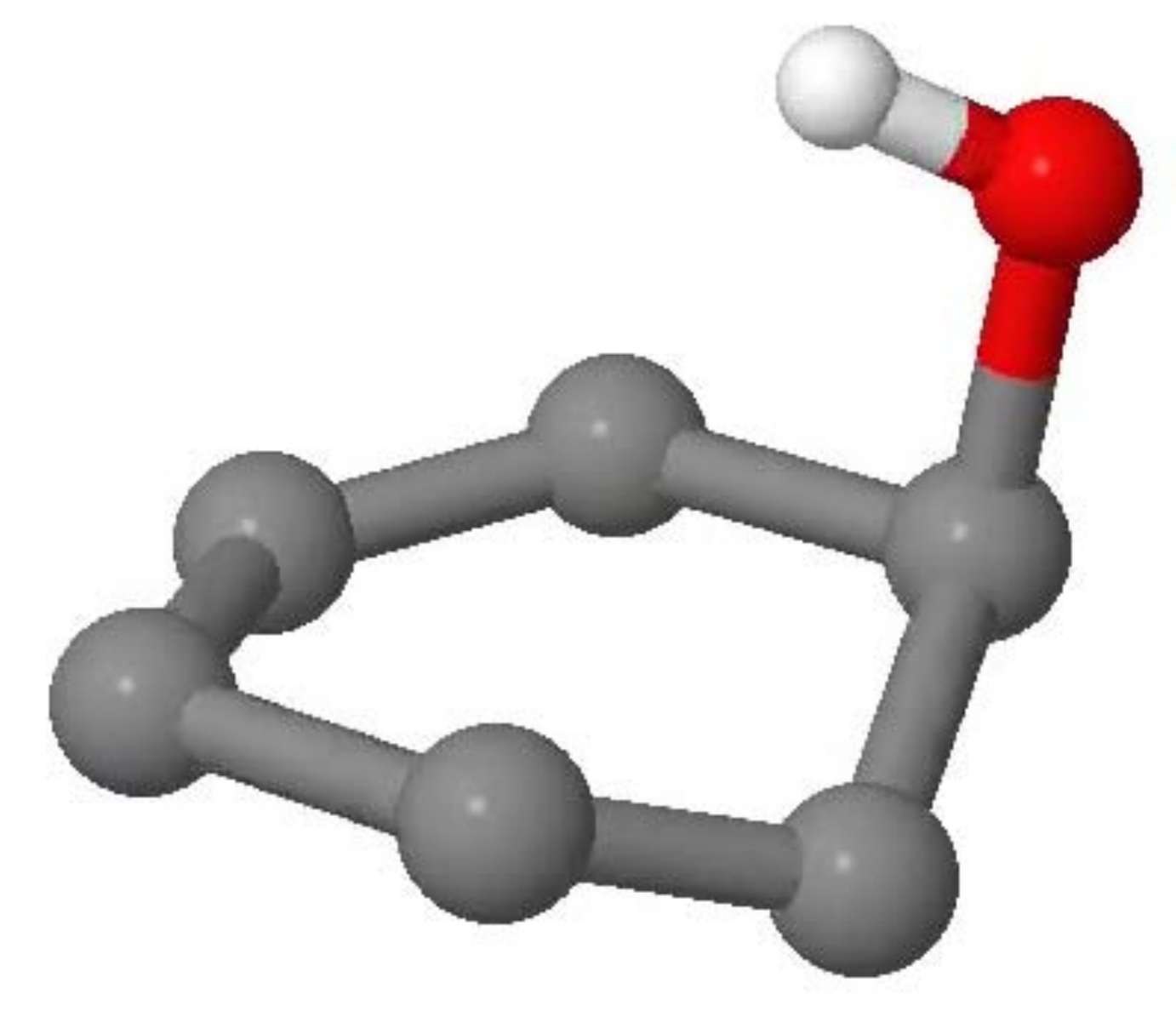}
\setfloatlink{https://nanoten.com/tomanek/manuscripts/DT290vA13.mp4} %
\caption{Hydroxyl group detachment from GO at the simulation
temperature
$2,500$~K.%
}
\label{videoA13}
\end{video}

Video~\ref{videoA13} illustrates the high stability of the bond
between a hydroxyl group and the carbon substrate. This bond
eventually breaks at 2,500~K in a process involving a low
activation barrier of 0.27~eV according to DFT
calculations~\cite{Lu2011}.

\begin{video}[h]
\includegraphics[width=0.50\columnwidth]{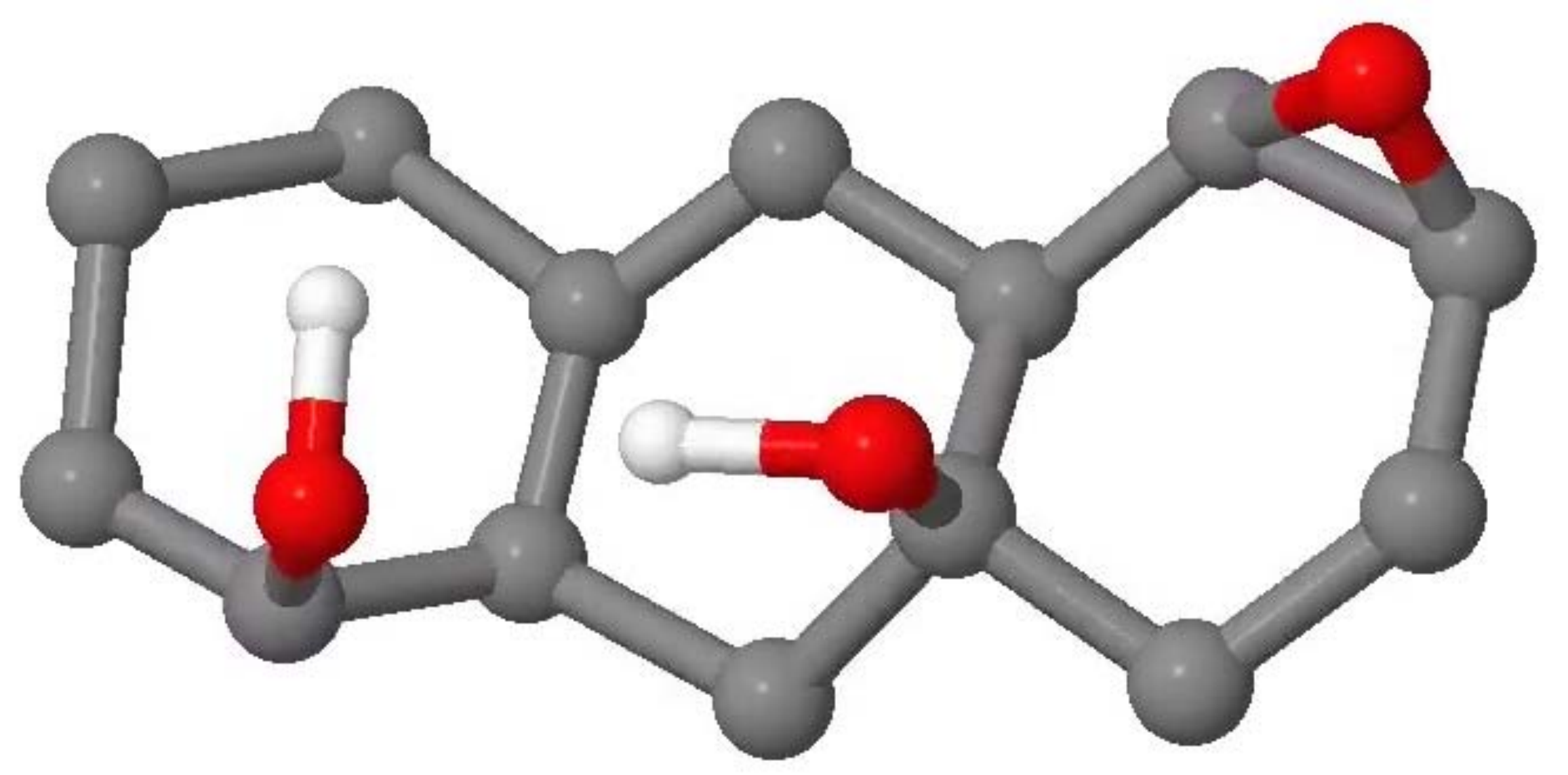}
\setfloatlink{https://nanoten.com/tomanek/manuscripts/DT290vA14.mp4} %
\caption{Chain reaction involving hydrogen transfers between epoxy
and OH-groups at the simulation temperature $2,500$~K.%
}
\label{videoA14}
\end{video}

Video~\ref{videoA14} shows how different types of
oxygen-containing groups may inter-convert by hydrogen exchange at
$2,500$~K. The lowest activation energy for such an exchange has
been calculated to be 0.18~eV for OH- and epoxy-groups in adjacent
1,2-sites, but to increase significantly to 0.88~eV for
1,3-sites~\cite{Lu2011}. Our simulation illustrates two exchange
events for 1,3-sites in a chain reaction involving hydrogen
transfer. This process starts with an OH-OH-epoxy arrangement.
After two exchange reactions, an epoxy-OH-OH configuration is
formed. Evidently, the activation barriers are easily overcome at
this high temperature, which results in frequent epoxy-OH
conversions.

\begin{video}[h]
\includegraphics[width=0.50\columnwidth]{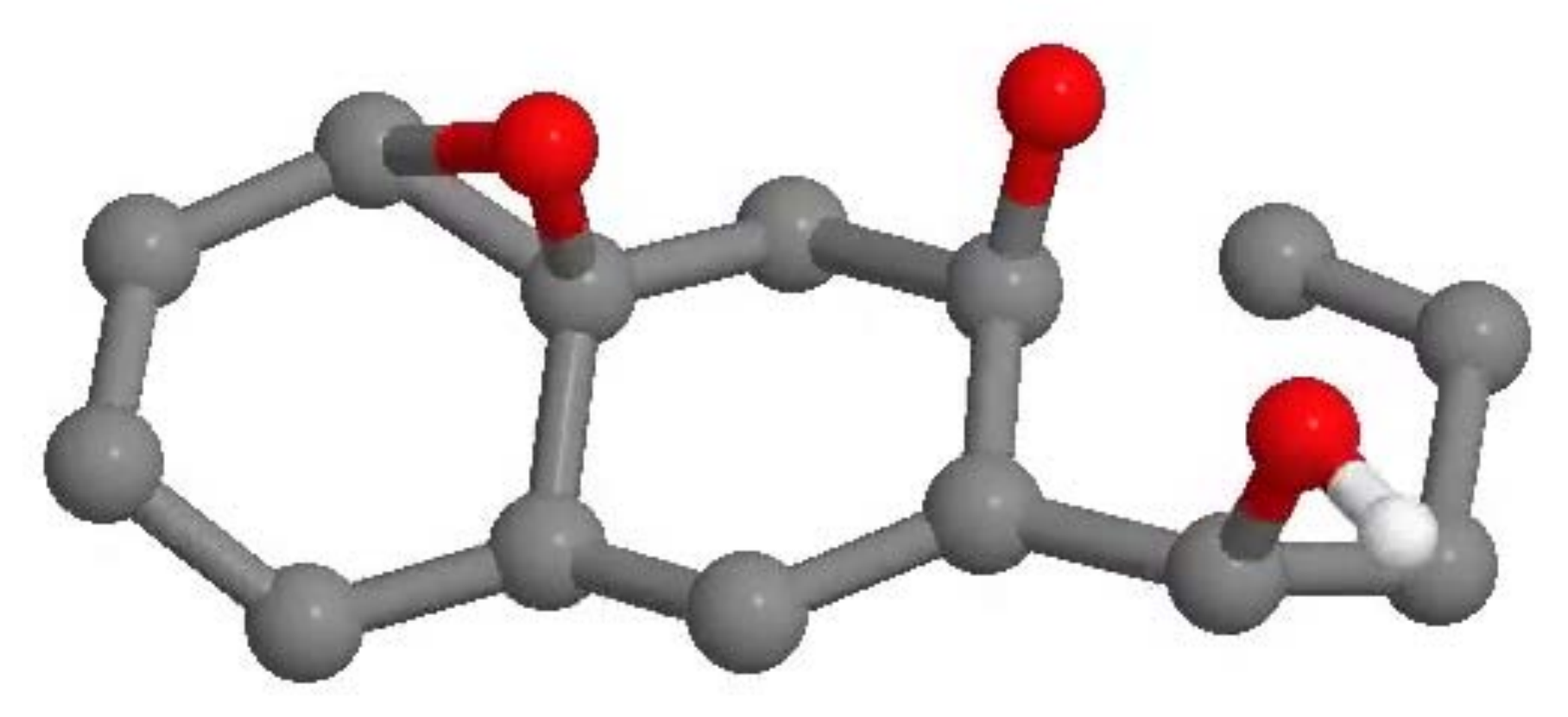}
\setfloatlink{https://nanoten.com/tomanek/manuscripts/DT290vA15.mp4} %
\caption{Hydrogen exchange followed by a C-O-O-C peroxy group
formation at the simulation temperature $2,500$~K.%
}
\label{videoA15}
\end{video}

As seen in Video~\ref{videoA15}, a similar process involving
hydrogen exchange reactions may lead to the formation of
mono-coordinated oxygen atoms instead of epoxy-groups, often
facilitated by the puckering of GO backbone at 2,500~K. In the
specific case visualized in the movie, an epoxy group and a nearby
mono-coordinated oxygen may convert to a C-O-O-C peroxide
structure. With its bond enthalpy below 2.07~eV~\cite{CRC-CP62},
the peroxide bond is very weak and breaks easily, as seen further
on in the video. Besides an intermittent C-C bond fracture near
the functional groups, we do not observe serious structural damage
to the GO backbone at 2,500~K.


\begin{acknowledgments}
We appreciate the information provided by Alexandr Talyzin about
processes observed in hydrated GO at elevated temperatures and the
assistance of Barbod Naderi with the interpretation of
atomic-level processes in the system. Computational resources for
this study have been provided by the Michigan State University
High Performance Computing Center.
\end{acknowledgments}

%
%
%



\begin{thebibliography}{65}
\bibitem [{MD-()}]{MD-temperature}%
  \BibitemOpen
  \href@noop {} {}\bibinfo {note} {The need for an artificial temperature
  increase in thermally accelerated MD simulations is discussed in Section
  II.}\BibitemShut {Stop}%
\end{thebibliography}

%

\end{document}